\documentclass[twocolumn,trackchanges]{aastex631}
\usepackage{multirow}
\usepackage{amsmath}
\usepackage{float}
\usepackage[normalem]{ulem}

\begin{document}

\title{Gamma-ray Orbital Modulation in Spider Pulsars:\\ Three Discoveries and a Universal Modulated Fraction}

\correspondingauthor{Maksat Satybaldiev}
\email{maksat.satybaldiev@ntnu.no}

\author[0009-0001-3795-4049]{Maksat Satybaldiev}
\affiliation{Department of Physics, Norwegian University of Science and Technology, NO-7491 Trondheim, Norway}
% \email{maksat.satybaldiev@ntnu.no}

\author[0000-0002-0237-1636]{Manuel Linares}
\affiliation{Department of Physics, Norwegian University of Science and Technology, NO-7491 Trondheim, Norway}
\affiliation{Departament de F{\'i}sica, EEBE, Universitat Polit{\`e}cnica de Catalunya, Av. Eduard Maristany 16, E-08019 Barcelona, Spain}
% \email{manuel.linares@ntnu.no}

\author[0000-0002-9802-3678]{Vittoria Vecchiotti}
\affiliation{Department of Physics, Norwegian University of Science and Technology, NO-7491 Trondheim, Norway}
\affiliation{Tsung-Dao Lee Institute, Shanghai Jiao Tong University, Shanghai 201210, P. R. China}
\affiliation{INAF Osservatorio Astrofisico di Arcetri, Largo Enrico Fermi, 5, 50125, Florence, Italy}
% \email{vittoria.vecchiotti@ntnu.no}



\begin{abstract}
Compact binary millisecond pulsars (also known as spiders) allow us to probe pulsar winds in their innermost regions, between the light cylinder (radius $\sim10^{7}$~cm) and the companion star (at $\sim10^{11}$~cm). Their flux is known to vary along the orbit, from radio to X-rays. 
During the past decade, gamma-ray orbital modulation (GOM) has been discovered in a handful of spiders, but its origin remains largely unknown. 
We present the results of a systematic search for GOM among 43 systems, selecting pulsed 0.1-1~GeV photons and using spin and orbital ephemeris from {\it Fermi}'s Third Pulsar Catalog.
We discover GOM from three spiders -- PSR~J1124-3653, PSR~J1946-5403 and PSR~J2215+5135 -- and confirm four previous detections.
In all seven cases so far, the GOM peaks near the pulsar's superior conjunction.
The X-ray orbital light curves are usually in antiphase, peaking when the pulsar is at inferior conjunction, but we find one case where both gamma-rays and X-rays peak around superior conjunction: PSR~J1946-5403.
We measure the modulated fractions of the GOM and find consistent values for all seven spiders, with an average $22.0\pm2.6\%$.
Including eclipsing systems seen edge-on, we find no clear dependence of the modulated fraction on the orbital inclination (within $\simeq$45-90$^\circ$). 
Our results challenge previous models proposed to explain GOM in spiders, based on inverse Compton and synchrotron emission close to the companion, since these predict a clear dependence on orbital inclination (stronger modulation at high inclinations). 
We nearly double the number of GOM detections in spiders, showing that it is more common than previously thought.
\end{abstract}



\section{Introduction} \label{sec:intro}

Compact binary millisecond pulsars or \textit{spiders} are binary systems consisting of a millisecond pulsar (MSP) and a low-mass  nondegenerate or semidegenerate star in a compact orbit with a period $P_\textrm{b}\lesssim1$ d.
The arachnid nickname was inspired by the destructive effect of the pulsar wind on the companion star. 
Spiders are categorized into two subgroups: \textit{redbacks} with companion masses $M_c\sim0.1M_\odot$ and \textit{ black widows}, which have significantly lighter companions $M_c\sim0.01M_\odot$ \citep{roberts13, koljonen25}.

Spiders represent the evolutionary link between low-mass X-ray binaries (LMXBs) and MSPs. 
During the LMXB phase, due to angular momentum transfer via accretion, the neutron star is spun up to millisecond periods \citep{alpar82, binovatyi76}. 
The so-called transitional MSPs switch between a disk or underluminous LMXB state and a rotation-powered redback MSP state, and provided key evidence for this evolutionary link \citep{archibald09, papitto13, bassa14}.
During the transition from the MSP to the LMXB state, the systems brighten across the optical, X-ray and gamma-ray bands, and the  pulsed radio emission disappears \citep{stappers13, patruno14, takata14, stappers14}.
{\it In the LMXB state}, the optical and X-ray emission is dominated by the accretion disk, and the X-ray flux switches between two distinct modes \citep{linares14b}.
The gamma-rays are believed to originate from the inverse Compton emission of the pulsar-wind-heated disk and from synchrotron radiation of pulsar wind particles heated by collisions with the disk \citep{veledina19}.

{\it In the MSP state}, spiders emit and exhibit orbital modulation across the entire electromagnetic spectrum.
In the radio band, we observe millisecond pulsations from the pulsar that are often occulted during a large fraction of the orbit around the pulsar's superior conjunction. 
These occultations (also referred to as ``eclipses") are produced by the scattering of pulsed radio emission from the pulsar by the ablated material from the companion \citep{fruchter88, thompson94, blanchard25}. 
At optical wavelengths, the emission primarily originates from the companion star and varies throughout the orbit due to irradiation from the pulsar wind and ellipsoidal modulation caused by tidal deformation of the star \citep{breton13, linares17, strader19, turchetta23}. 

The X-ray emission in the MSP state consists of thermal and nonthermal components. 
In a few cases, thermal emission is detected from the heated polar caps, and can be pulsed \citep{guillot19}.
The typically dominant non-thermal component is thought to arise from an intrabinary shock (IBS) formed by the interaction between the pulsar and companion winds \citep{wadiasingh17,wadiasingh18, vandermerwe20}.
The X-ray orbital light curve exhibits a distinct single- or double-peaked structure, attributed to Doppler-boosted synchrotron emission from particles accelerated in the shock.
The shock geometry depends on the relative strengths of the winds: it can wrap around the pulsar if the companion's wind has the greater momentum, 
or around the companion in the opposite case.
When the shock is wrapped around the pulsar, the maximum of the X-ray orbital modulation (XOM) is expected near the pulsar's inferior conjunction, when the beamed IBS emission crosses the line of sight \citep{wadiasingh17,wadiasingh18, vandermerwe20}.

\begin{deluxetable}{llllll}
\tablewidth{\textwidth}
\tabletypesize{\scriptsize}
\tablecaption{Spider pulsars with significant gamma-ray orbital modulation (GOM) reported. See text for details.
\label{tab:detected_gom}}
\tablehead{
Name & Type & GOM & GOM & XOM & OOM\\
     &   & type & peak & peak & 
}
\startdata
\textbf{J1124-3653} & BW & Soft, pulsed & SC [1] & IC [2,3] & 1 [4]\\
J1227-4853 & RBt & Soft & SC [5] & IC [6,7] & 1 [8]\\
\textbf{J1946-5403} & BW & Soft, pulsed & SC [1] & SC [3] & * [9] \\
J2039-5617 & RB & Soft, pulsed & SC [10,11] & IC [12,13] & 2 [8] \\
\textbf{J2215+5135} & RB & Soft, pulsed & SC [1] & IC [14]& 1 [8] \\
J2241-5236 & BW & Soft, pulsed & SC [15] &  & 1 [8]\\
J2339-0533 & RB & Soft, pulsed & SC [16] & IC [17,18] & 1 [8]\\
\hline
J1311-3430 & BW & Hard, off-pulse & IC [19,20] & & 1 [8]\\
J1702.7-5655 & RBc & Full & IC [21] & &  \\
J1023+0038 ** & RBt & Hard & DN [22] & IC [23] & 1 [8]\\
\hline
\enddata
\tablecomments{RB: redback; RBt: redback and transitional MSP; RBc: redback candidate; BW: black widow; SC: pulsar's superior conjunction; IC: pulsar's inferior conjunction; DN: descending node of the puslar; OOM: number of maxima in the optical orbital modulation - 1(irradiated light curve) or 2 (ellipsoidal modulation); *: No optical counterpart confirmed for J1946, but see search in [9]; ** - for J1023+0038, the GOM corresponds to the disk state, whereas the XOM is observed in the pulsar state.\\
Systems with GOM discovered in this paper are highlighted in bold.\\
References: [1] - this work, [2] - \citealt{gentile14}, [3] - \citealt{sim2024}, [4] - \citealt{draghis19}, [5] - \citealt{an22}, [6] - \citealt{demartino15}, [7] - \citealt{demartino20}, [8] - \citealt{turchetta23}, [9] - \citealt{braglia20}, [10] -\citealt{ng18}, [11] - \citealt{clark21}, [12] - \citealt{salvetti15}, [13] - \citealt{romani15}, [14] - \citealt{sullivan25}, [15] - \citealt{an18}, [16] - \citealt{an20}, [17] - \citealt{romani11}, [18] - \citealt{kandel19}, [19] \citealt{xing15}, [20] - \citealt{an17}, [21] - \citealt{corbet22}, [22] - \citealt{xing18}, [23] - \citealt{bogdanov11}.
}
% \tablenotetext{*}{text}
\end{deluxetable}
 
Pulsed gamma-ray emission in neutron stars originates in the current sheet just outside the light cylinder and peaks at GeV energies \citep{philippov22, hakobyan23, 3PC, cerutti25}.
The pulse-averaged gamma-ray flux in isolated pulsars is stable within $<10\%$ \citep{kerr25}.
Some of the known spider pulsars exhibit eclipses, showing sharp dips in their gamma-ray light curves \citep{corbet22, clark23}. 
In addition, the gamma-ray flux can also be orbitally modulated.
Orbital modulation of the {\it soft} gamma-ray flux ($\lesssim 1$ GeV) has been reported in four systems:
the redbacks PSR J2039-5617 and PSR J2339-5617, the black widow PSR J2241-5236 and the transitional MSP PSR J1227-4853 (see references in Table~\ref{tab:detected_gom}).
In the first two systems, the gamma-ray orbital modulation (GOM) was found to be pulsed, i.e., arising from photons corresponding to the peak of the pulse profile.
In all four cases, the soft GOM peaks at the pulsar's superior conjunction.
Interestingly, in systems where orbital modulation is observed in both X-ray and gamma-ray bands, the orbital light curves are in antiphase: the gamma-ray flux peaks near orbital phase $\sim0.25$ (pulsar's superior conjunction), while the X-ray flux peaks near phase $\sim0.75$ (pulsar's inferior conjunction) \citep[see, e.g.,][]{sim24}.

Different types of GOM have been observed in three other spider systems. 
The black widow PSR J1311-3430 shows off-pulse {\it hard} orbital modulation (in the $1-100$ GeV band), peaking at orbital phase $\Phi\approx0.8$, following the pulsar's inferior conjunction \citep{xing15, an17}.
\cite{corbet22} reported gamma-ray eclipses and modulation in the candidate redback system 4FGL J1702.7-5655, which peaked about half an orbit after the eclipse and was detected only after MJD $\sim56345$.
In addition, \citealt{xing18} reported a possible GOM at higher energies ($>5.5$ GeV) in the disk state of the transitional MSP PSR J1023+0038, peaking around the descending node of the pulsar.
All systems with reported GOM are summarized in Table \ref{tab:detected_gom}, where we also list information about their XOM and optical orbital modulation. 
In this work, we focus on searching for and studying {\it soft and pulsed GOM}.

While the IBS synchrotron emission is widely accepted as the source of XOM, the origin of GOM in spiders remains 
a matter of debate.
Boosted synchrotron emission from electrons along the IBS cannot account for the half-orbit phase shift of the gamma-ray light curve. 
The first explanation explored by \citet{an20} involves gamma-ray flux suppression by the swept-back pulsar wind. 
In this scenario, the intrinsically constant pulsar emission is attenuated at low energies due to orbital phase dependent absorption by the electrons in the pulsar wind.
Scattering in the Klein-Nishina regime could produce the modulation in the soft gamma-ray band. 
However, the optical depth achievable within the system was shown to be insufficient to account for the observed amplitude of the GOM \citep{an20}, so we do not discuss this model below.
Since then, two main models have been proposed to explain the GOM in spiders, namely: 

(i) additional flux from inverse Compton (IC) scattering of companion blackbody photons by pulsar wind particles and IBS electrons \citep{ng18, an20, clark21, sim24}. The emission geometry in this scenario naturally explains the observed phasing of the modulation. However, the predicted IC flux is insufficient to match the observations. Matching the observed flux would require the deceleration of the upstream pulsar wind, which would increase the residence time of the electrons and boost the modulated emission. However, such deceleration would suppress or even prevent the formation of the IBS, which is necessary to explain the X-ray data \citep{sim24}. 

(ii) synchrotron radiation from relativistic leptons ejected by the pulsar interacting with the companion's magnetic field \citep{vandermerwe20, clark21, sim24}. This scenario can reproduce both the observed fluxes and the phasing of the modulation, provided the leptons are accelerated to energies of $\sim0.1$ PeV and the companion's magnetic field is $\sim 1$ kG. These conditions are also consistent with the presence of an IBS, allowing it to form and emit as required by the X-ray data.

In this work, we present the results of a systematic search for soft and pulsed GOM from confirmed spider systems using the Third Fermi Large Area Telescope (LAT) Catalog of Gamma-Ray Pulsars \cite[3PC,][]{3PC}.
In Section \ref{sec:analysis}, we describe the pulsar sample and analysis methods.
We present the results of the search in Section \ref{sec:results}, including the discovery of GOM in three spider systems.
Finally, in Section \ref{sec:discussion} we discuss the properties of soft pulsed GOM and compare them with other system parameters, in the context of the proposed models.

\section{Observations and analysis} \label{sec:analysis}

\subsection{Spider pulsar sample}
\label{sec:sample}

Out of 294 gamma-ray pulsars reported in 3PC, we preselected all 47 systems that are also listed as confirmed spiders in the SpiderCat catalog of compact binary MSPs\footnote{Version 1.8.1, at \url{https://astro.phys.ntnu.no/SpiderCAT}} \citep[][]{koljonen25}: 15 redbacks and 32 black widows.
Three black widows (PSR J1311-3430, PSR J1745+1017, PSR J2115+5448) and one redback (PSR J0955-3947) in this list had diluted precision timing solutions, pending publications.
Since sensitive GOM searches require precise spin and orbital ephemerides, these four systems were excluded from our analysis, leaving a total of 43 systems selected for further investigation (14 redbacks and 29 black widows).
For the same reason, we did not include spider candidates, although GOM may be present in some of these systems as well \citep{ng18, corbet22}.
The final list of studied pulsars and their main properties is presented in Table \ref{tab:properties}.

\begin{deluxetable*}{lcccccccccc}
\tablewidth{\textwidth}
\setlength{\tabcolsep}{2pt}
\tablecaption{Main properties of the 43 spider pulsars studied here, from the 4FGL-DR4 \citep{abdollahi22, ballet23} and 3PC \citep{3PC} catalogs: 4FGL counterpart name, spin period $P_\textrm{s}$, spin period derivative $\dot{P}$, orbital period $P_\textrm{b}$, phase-averaged $0.1-300$ GeV energy flux $F_\textrm{g}$, estimated distance $d$, spin-down power $\dot{E}$, gamma-ray luminosity in the $0.1-100$ GeV energy band $L_\textrm{g}$ (the second set of uncertainties is due to the distance) and projected semimajor axis $A1$. In the name column (e) indicates gamma-ray eclipsing binaries, while (t) - transitional systems.
\label{tab:properties}}
\tablehead{
% \multirow{2}{*}{Name} &
% \multirow{2}{*}{Type} & 
% \multirow{2}{*}{4FGL name} & 
\colhead{Name} &
\colhead{Type} & 
\colhead{4FGL name} & 
\colhead{$P_\textrm{s}$} & 
\colhead{$\dot{P}/10^{-20}$} &
\colhead{$P_\textrm{b}$} & 
\colhead{$F_\textrm{g}/10^{-12}$} & 
\colhead{$d$} &  
\colhead{$\dot{E}/10^{33}$} & 
\colhead{$L_\textrm{g}/10^{33}$} & 
\colhead{A1}\\
\colhead{} & 
\colhead{} & 
\colhead{} & 
\colhead{ms} & 
\colhead{s\ s$^{-1}$} & 
\colhead{days} & 
\colhead{erg cm$^{-2}$ s$^{-1}$} & 
\colhead{kpc} &  
\colhead{erg s$^{-1}$} & 
\colhead{erg s$^{-1}$} & 
\colhead{lt-s}
}
\startdata
J0023+0923 & BW & J0023.4+0920 & 3.05 & 1.14 & 0.14 & $7.8\pm0.6$ & $1.8^{+0.5}_{-0.3}$ & $12.4^{+1.0}_{-0.7}$ & $3.03\pm0.21^{+1.97}_{-1.00}$ & 0.035\\
J0251+2606 & BW & J0251.0+2605 & 2.54 & 0.76 & 0.20 & $4.9\pm0.4$ & $1.2\pm0.5$ & $12.8\pm3.2$ & $0.80\pm0.07^{+0.77}_{-0.51}$ & 0.066\\
J0312-0921 & BW & J0312.1-0921 & 3.70 & 1.97 & 0.10 & $5.5\pm0.4$ & $0.82\pm0.33$ & $9.7\pm2.4$ & $0.44\pm0.03^{+0.43}_{-0.28}$ & 0.015\\
J0610-2100 & BW & J0610.2-2100 & 3.86 & 1.23 & 0.29 & $7.2\pm0.5$ & $2.2\pm0.7$ & $3.5\pm1.6$ & $4.19\pm0.27^{+3.09}_{-2.24}$ & 0.073\\
J0636+5128 & BW & - & 2.87 & 0.34 & 0.07 & - & $0.7^{+0.2}_{-0.1}$ & $5.60^{+0.06}_{-0.04}$ & - & 0.009\\
J0952-0607 & BW & J0952.1-0607 & 1.41 & 0.48 & 0.27 & $2.4\pm0.3$ & $6.3^{+0.3}_{-0.4}$ & $\leq 68.5$ & $11.04\pm1.47^{+1.08}_{-1.37}$ & 0.063\\
J1023+0038 (t) & RB & J1023.7+0038 & 1.69 & 0.68 & 0.20 & $32.4\pm0.8$ & $1.37\pm0.04$ & $42.8\pm0.4$ & $7.25\pm0.18\pm0.43$ & 0.343\\
\multicolumn{1}{l}{Pulsar state} &  &  &  &  & $5.6\pm1.2$ & & & & \\
\multicolumn{1}{l}{Disk state} &  &  &  &  & $49.2\pm0.7$ & & & & \\
J1048+2339 (e) & RB & J1048.6+2340 & 4.67 & 3.01 & 0.25 & $4.9\pm0.5$ & $1.3^{+0.6}_{-0.2}$ & $9.7^{+0.9}_{-0.4}$ & $1.02\pm0.10^{+1.05}_{-0.32}$ & 0.836\\
J1124-3653 & BW & J1124.0-3653 & 2.41 & 0.60 & 0.23 & $12.5\pm0.6$ & $1.0\pm0.4$ & $16.7\pm0.3$ & $1.46\pm0.07^{+1.41}_{-0.94}$ & 0.080\\
J1221-0633 & BW & J1221.4-0634 & 1.93 & 0.53 & 0.39 & $5.8\pm0.5$ & $1.3\pm0.5$ & $> -165$ & $1.09\pm0.09^{+1.05}_{-0.70}$ & 0.055\\
J1227-4853 (t) & RB & J1228.0-4853 & 1.69 & 1.33 & 0.29 & $19.0\pm1.6$ & $1.8^{+0.5}_{-0.2}$ & $87.3^{+6.7}_{-2.8}$ & $6.98\pm0.60^{+4.46}_{-1.50}$ & 0.668\\
\multicolumn{1}{l}{Pulsar state} &  &  &  &  & $16.3\pm2.8$ & & & & \\
\multicolumn{1}{l}{Disk state} &  &  &  &  & $32.0\pm2.2$ & & & & \\
J1301+0833 & BW & J1301.6+0834 & 1.84 & 1.06 & 0.27 & $7.7\pm0.5$ & $1.8\pm0.1$ & $43.7\pm8.2$ & $2.88\pm0.18\pm0.33$ & 0.078\\
J1302-3258 & RB & J1302.4-3258 & 3.77 & 0.65 & 0.78 & $10.9\pm0.5$ & $1.4\pm0.6$ & $3.6\pm0.7$ & $2.67\pm0.13^{+2.56}_{-1.71}$ & 0.928\\
J1431-4715 & RB & J1431.4-4711 & 2.01 & 1.41 & 0.45 & $4.7\pm0.7$ & $1.8^{+0.5}_{-0.2}$ & $54.3^{+4.5}_{-1.8}$ & $1.74\pm0.26^{+1.20}_{-0.39}$ & 0.550\\
J1446-4701 & BW & J1446.6-4701 & 2.19 & 0.98 & 0.28 & $7.7\pm0.7$ & $1.6\pm0.6$ & $36.2\pm0.4$ & $2.26\pm0.20^{+2.17}_{-1.45}$ & 0.064\\
J1513-2550 & BW & J1513.4-2549 & 2.12 & 2.15 & 0.18 & $7.6\pm0.6$ & $4.0\pm1.6$ & $86.8^{+2.7}_{-2.1}$ & $14.27\pm1.16^{+13.70}_{-9.13}$ & 0.041\\
% J1526-2744 & None & J1526.6-2743 & 2.49 & 0.20 & $2.6\pm0.4$ & $1.3\pm0.5$ & $\leq 9.2$ & $0.54\pm0.09^{+0.52}_{-0.35}$ & 0.224\\
J1544+4937 & BW & J1544.0+4939 & 2.16 & 0.28 & 0.12 & $2.4\pm0.3$ & $3.0^{+2.0}_{-1.0}$ & $8.4^{+4.4}_{-2.2}$ & $2.53\pm0.31^{+4.53}_{-1.40}$ & 0.033\\
J1555-2908 (e) & BW & J1555.7-2908 & 1.79 & 4.45 & 0.23 & $4.7\pm0.6$ & $5.1^{+0.5}_{-0.7}$ & $> 306$ & $14.51\pm1.88^{+2.98}_{-3.71}$ & 0.151\\
J1622-0315 & RB & J1623.0-0315 & 3.85 & 1.14 & 0.16 & $7.2\pm0.7$ & $1.2^{+0.5}_{-0.2}$ & $6.6^{+0.6}_{-0.2}$ & $1.31\pm0.12^{+1.40}_{-0.36}$ & 0.219\\
J1627+3219 & BW & J1627.7+3219 & 2.18 & 0.55 & 0.17 & $3.6\pm0.3$ & $4.5\pm1.8$ & $\leq 24.8$ & $8.71\pm0.75^{+8.36}_{-5.58}$ & 0.054\\
J1628-3205 & RB & J1628.1-3204 & 3.21 & 1.19 & 0.21 & $11.3\pm0.9$ & $1.2^{+0.6}_{-0.2}$ & $11.9^{+2.9}_{-0.9}$ & $1.92\pm0.16^{+2.53}_{-0.57}$ & 0.410\\
J1641+8049 & BW & J1641.2+8049 & 2.02 & 0.98 & 0.09 & $2.0\pm0.3$ & $3.0\pm1.2$ & $> 48$ & $2.19\pm0.34^{+2.10}_{-1.40}$ & 0.064\\
J1653-0158 & BW & J1653.6-0158 & 1.97 & 0.24 & 0.05 & $34.3\pm1.0$ & $0.7^{+1.6}_{-0.1}$ & $7.0^{+12.0}_{-0.8}$ & $2.25\pm0.07^{+19.48}_{-0.35}$ & 0.011\\
J1803-6707 & BW & J1803.1-6708 & 2.13 & 1.85 & 0.38 & $4.8\pm0.5$ & $1.8^{+0.8}_{-0.3}$ & $71.7^{+2.0}_{-0.7}$ & $1.94\pm0.20^{+2.11}_{-0.56}$ & 1.062\\
J1805+0615 & BW & J1805.6+0615 & 2.13 & 2.28 & 0.34 & $5.3\pm0.5$ & $3.9\pm1.6$ & $75.9^{+10.7}_{-10.5}$ & $9.51\pm0.98^{+9.13}_{-6.09}$ & 0.088\\
J1810+1744 & BW & J1810.5+1744 & 1.66 & 0.45 & 0.15 & $23.3\pm0.9$ & $1.9^{+1.1}_{-0.4}$ & $36.6^{+3.3}_{-1.2}$ & $9.73\pm0.39^{+15.14}_{-3.55}$ & 0.095\\
J1816+4510 (e) & RB & J1816.5+4510 & 3.19 & 4.31 & 0.36 & $10.6\pm0.5$ & $3.4^{+1.2}_{-0.5}$ & $53.1^{+0.4}_{-0.2}$ & $14.88\pm0.70^{+12.35}_{-3.87}$ & 0.595\\
J1833-3840 & BW & J1833.0-3840 & 1.87 & 1.77 & 0.90 & $2.8\pm0.4$ & $4.6\pm1.9$ & $\leq 105.1$ & $7.29\pm1.16^{+7.00}_{-4.66}$ & 0.061\\
J1908+2105 & RB & J1908.9+2103 & 2.56 & 1.38 & 0.15 & $4.9\pm0.8$ & $2.6\pm1.0$ & $30.9^{+2.0}_{-1.9}$ & $3.95\pm0.64^{+3.79}_{-2.53}$ & 0.117\\
J1946-5403 & BW & J1946.5-5402 & 2.71 & 0.27 & 0.13 & $9.8\pm0.5$ & $1.1\pm0.5$ & $> 5$ & $1.56\pm0.08^{+1.49}_{-1.00}$ & 0.044\\
B1957+20 (e) & BW & J1959.5+2048 & 1.61 & 1.68 & 0.38 & $15.7\pm0.9$ & $1.3^{+0.8}_{-0.3}$ & $116.9^{+25.7}_{-9.4}$ & $3.02\pm0.18^{+4.63}_{-1.15}$ & 0.089\\
J2017-1614 & BW & J2017.7-1612 & 2.31 & 0.24 & 0.10 & $6.5\pm0.6$ & $1.4\pm0.6$ & $> -26$ & $1.61\pm0.15^{+1.55}_{-1.03}$ & 0.044\\
J2039-5617 & RB & J2039.5-5617 & 2.65 & 1.42 & 0.23 & $15.4\pm0.6$ & $1.7^{+0.6}_{-0.2}$ & $25.4^{+1.8}_{-0.8}$ & $5.21\pm0.22^{+3.97}_{-1.28}$ & 0.471\\
J2047+1053 & BW & J2047.3+1051 & 4.29 & 2.08 & 0.12 & $4.3\pm0.6$ & $2.8\pm1.1$ & $6.8\pm1.9$ & $3.98\pm0.52^{+3.82}_{-2.55}$ & 0.069\\
J2051-0827 & BW & J2051.0-0826 & 4.51 & 1.27 & 0.10 & $2.5\pm0.3$ & $1.47\pm0.59$ & $5.5\pm0.1$ & $0.65\pm0.08^{+0.62}_{-0.41}$ & 0.045\\
J2052+1219 & BW & J2052.7+1218 & 1.99 & 0.67 & 0.11 & $4.6\pm0.6$ & $5.6^{+0.5}_{-0.6}$ & $10.5^{+3.4}_{-3.7}$ & $16.97\pm2.08^{+2.98}_{-3.19}$ & 0.061\\
J2129-0429 (e) & RB & J2129.8-0428 & 7.61 & 32.78 & 0.64 & $6.8\pm0.5$ & $1.8\pm0.1$ & $20.61\pm0.05$ & $2.73\pm0.19\pm0.34$ & 1.852\\
J2214+3000 & BW & J2214.6+3000 & 3.12 & 1.47 & 0.42 & $32.6\pm0.7$ & $0.6\pm0.3$ & $17.2\pm1.4$ & $1.41\pm0.03^{+1.83}_{-1.08}$ & 0.059\\
J2215+5135 & RB & J2215.6+5135 & 2.61 & 2.82 & 0.17 & $18.0\pm0.8$ & $2.7^{+1.5}_{-0.5}$ & $52.7^{+0.3}_{-0.1}$ & $15.26\pm0.69^{+21.16}_{-4.91}$ & 0.468\\
J2234+0944 & BW & J2234.7+0943 & 3.63 & 2.01 & 0.42 & $10.0\pm0.6$ & $0.7^{+0.2}_{-0.1}$ & $11.1^{+1.3}_{-0.9}$ & $0.59\pm0.04^{+0.31}_{-0.17}$ & 0.068\\
J2241-5236 & BW & J2241.7-5236 & 2.19 & 0.69 & 0.15 & $25.0\pm1.1$ & $1.04\pm0.05$ & $26.3^{+0.4}_{-0.3}$ & $3.25\pm0.14\pm0.29$ & 0.026\\
J2256-1024 & BW & J2256.8-1024 & 2.29 & 1.13 & 0.21 & $8.2\pm0.5$ & $2.1^{+0.9}_{-0.5}$ & $38.3^{+0.8}_{-0.4}$ & $4.25\pm0.26^{+4.74}_{-1.78}$ & 0.083\\
J2339-0533 & RB & J2339.6-0533 & 2.88 & 1.41 & 0.19 & $29.2\pm0.8$ & $1.3^{+0.3}_{-0.2}$ & $22.5^{+0.5}_{-0.3}$ & $5.90\pm0.15^{+3.61}_{-1.36}$ & 0.612\\
% \hline
\enddata
\end{deluxetable*}

All 43 spiders in our sample have counterparts in the Fermi-LAT point-source catalog \cite[4FGL,][]{abdollahi20, abdollahi22, ballet23}, with one exception: PSR~J0636+5128.
This pulsar is too faint to be included in the 4FGL catalog, 
which did not use pulsation searches.
However, when the data are phase-folded using a precise ephemeris, the pulsar becomes detectable \citep{3PC}.

\subsection{Fermi-LAT data analysis}
\label{sec:LAT}

To streamline our analysis,
we used the public data products from 3PC\footnote{\url{https://fermi.gsfc.nasa.gov/ssc/data/access/lat/3rd_PSR_catalog/}} \citep{3PC}: phased and weighted FT1 event files and timing solutions (\texttt{.par} files).
We selected \texttt{SOURCE} class photons with zenith angles $\leq90^\circ$ from a $3^\circ$-radius region of interest (RoI) around each source, covering two different energy ranges: soft $0.1-1$ GeV and full $0.1-300$ GeV within the period of validity of the timing solution.
For each photon, spin and orbital phases were calculated based on the timing model using \texttt{TEMPO2} \citep{tempo2, hobbs06} with the \texttt{Fermi} plugin \citep{tempo2fermi}.

Folded light curves were constructed using spin and orbital phases and probability weights from the \texttt{MODEL\_WEIGHT} column.
These weights represent the probability that a detected gamma-ray photon was emitted by the pulsar, based on the photon energy, arrival direction, and source model within the RoI.
Uncertainties for the $i$-th light curve bin are calculated as $\sigma_i^2 = \sum_{j=1}^{N_i} w_j^2 + (\max_{j\in[1,N_j]}w_j)^2$, where $N_i$ is the number of photons in the bin and $w_j$ are the photon weights.
The background level was estimated from the weights as $(\sum_{i}w_i-\sum_iw_i^2)/N_{bin}$ with an uncertainty of  $\pm6\%$ \citep{3PC}, where $w_i$ is the weight of the $i$-th photon and $N_{bin}$ is the number of phase bins in the light curve.
The background primarily arises from Galactic diffuse emission, the isotropic gamma-ray background, and contributions from nearby sources.
The significance of detected pulsations was determined using the standard weighted $H$-test with $m=20$ harmonics \citep{deJager89, deJager10, kerr11}.
This led to a significance of the detected pulsations of $>5\sigma$.

\begin{figure}[ht!]
\includegraphics[width=0.5\textwidth]{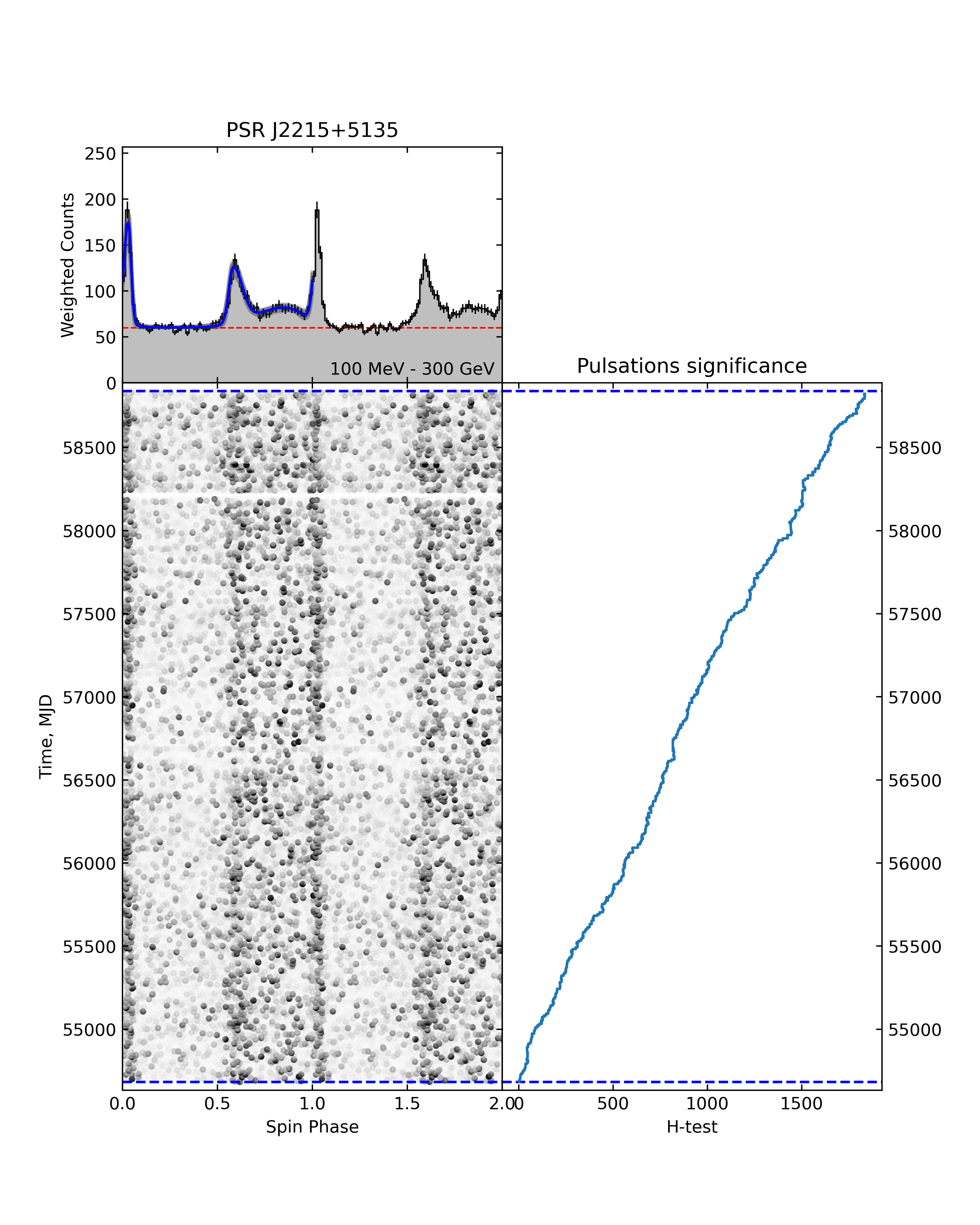}
\caption{Our gamma-ray timing results for PSR J2215+5135 in the $0.1-300$ GeV energy band. 
\textit{Top panel}: the black curve represents the folded light curve, the blue curve shows the template pulse profile, and the dashed red line indicates the background level. 
\textit{Bottom left}: photon spin phases are plotted against time, with the color representing the weights of the photons. 
\textit{Bottom right}: the cumulative H-test for the significance of the pulsations is shown over time. 
Dashed blue lines in the bottom panels indicate the start and end of the timing solution used. 
\label{fig:timing}}
\end{figure}

An example of the timing results is shown in Figure \ref{fig:timing}, which displays the pulse profile of PSR J2215+5135, the distribution of photon rotational phases over time (aligned using the 3PC ephemeris) and the H-test significance of the detected pulsations, which increases monotonically as more data are accumulated.
In this example, the precise phase alignment is achieved by modeling orbital period variations using Gaussian processes (ORBIFUNCS, see \citealt{clark21}).
In other cases, the orbital period variations are modeled with a Taylor series expansion around the time of ascending node (using the BTX or ELL1 models).

The weighted H-test was also used to quantify the significance of GOM using the calculated orbital phases.
To describe and compare the folded orbital light curves across different systems, we fitted them with a sinusoidal function by maximizing the likelihood
\begin{equation*}
\log\mathcal{L}=\sum_{j}\log\left[w_jF(\Phi_j)+(1-w_j)\right],  
\end{equation*}
where $F(\Phi)=\alpha\sin\left(2\pi\left(\Phi+\Phi_0\right)\right)+1$ approximates the orbital light curve, $\Phi$ is the orbital phase ($\Phi=0$ corresponds to the pulsars' ascending node), and $\Phi_0$ quantifies the phase shift of the maximum relative to phase 0.25.
Thus, the peak phase is given by $\Phi_{max}=0.25-\Phi_0$.
Although some orbital light curves are not strictly sinusoidal, we use this approach to quantify the amplitude and phase-centering of the GOM in an homogeneous way.
In this model, the parameter $\alpha$ is the semiamplitude of the modulation divided by the difference between the average flux and the background level; it provides a measure of the fractional semiamplitude of the orbital modulation, or modulated fraction (MF) hereafter.

\begin{figure}[ht!]
\includegraphics[width=0.45\textwidth]{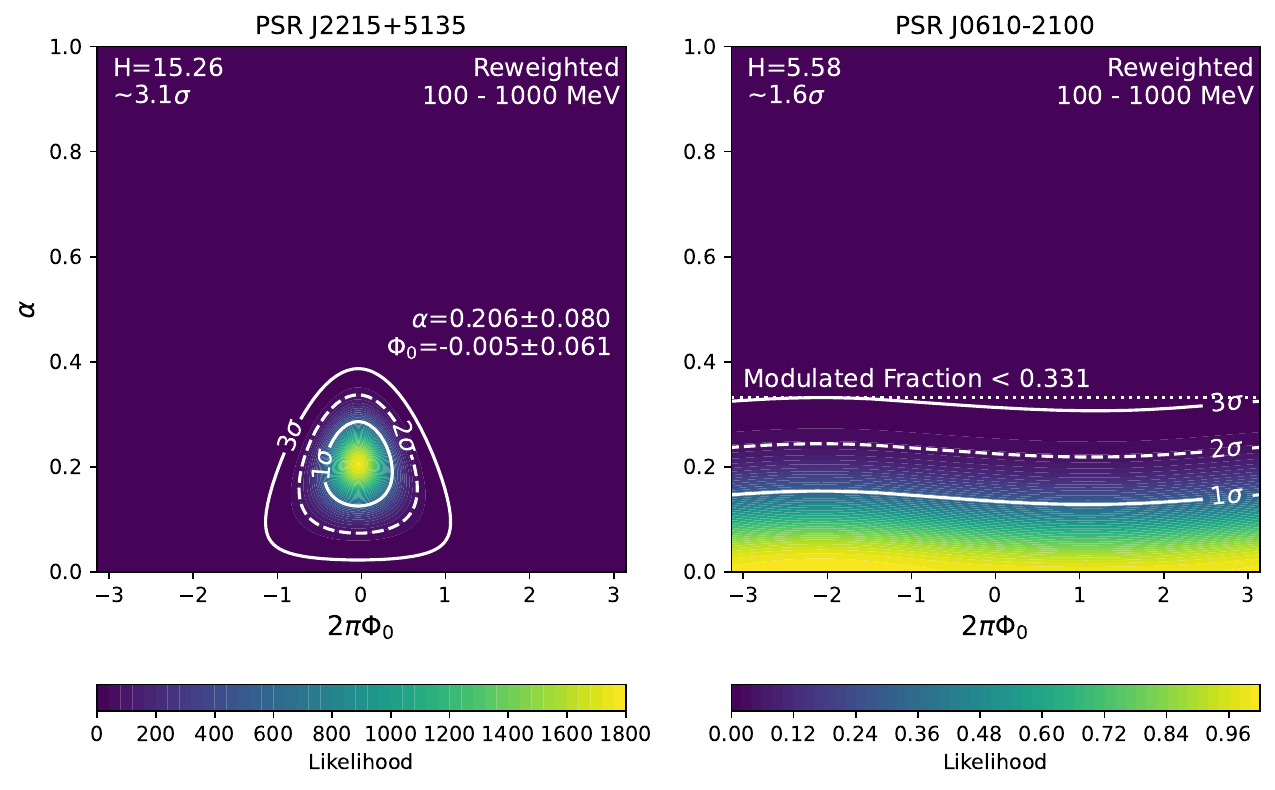}
\caption{
Constraints on the MF ($\alpha$) and phase shift ($\Phi_0$).
\textit{Left}: parameter space for PSR J2215+5135, with the color scale representing the likelihood values. Contours indicate the 1$\sigma$, 2$\sigma$, and 3$\sigma$ confidence intervals.
\textit{Right}: parameter space for PSR J0610-2100, which does not exhibit GOM. The 3$\sigma$ upper limits on the MF are shown.
\label{fig:par}}
\end{figure}

\begin{figure*}[ht!]
\includegraphics[width=1.0\textwidth]{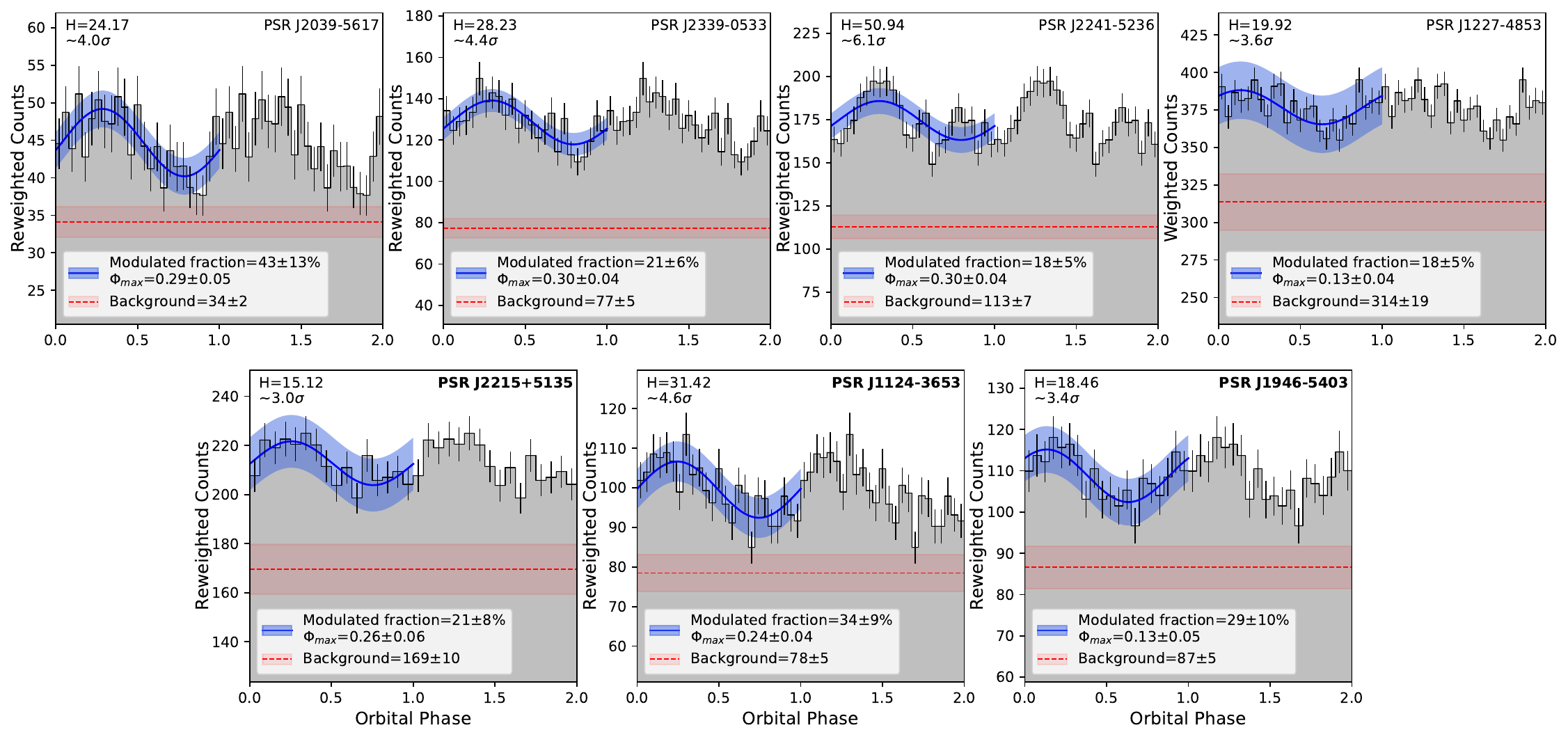}
\caption{Reweighted, phase-folded gamma-ray light curves in the $0.1–1$ GeV energy range for all systems with detected GOM (except for PSR J1227-4853, where regular weights were used; Secs.~\ref{sec:LAT} and \ref{sec:J1227}).
The black points represent the folded data and uncertainties. The blue line and shaded area show the sinusoidal fit with its $1\sigma$ uncertainty, while the red dashed line and shaded area represent the background level and its uncertainty. Two orbital periods are shown for clarity.
\label{fig:lcs}}
\end{figure*}

The confidence intervals for the model parameters, $\alpha$ and $\Phi_0$, were determined using Wilks' theorem
(i.e., assuming that the drop in log-likelihood, $-2\Delta\log\mathcal{L}$, follows a $\chi^2$ distribution with two degrees of freedom).
For sources with $H<14$ (corresponding to $<3\sigma$ significance), we placed $3\sigma$ upper limits on the MF. 
Figure \ref{fig:par} shows two such examples of how we determined uncertainties (left) and placed upper limits (right) on the fitted parameters.

From previous studies it was found that the soft GOM is pulsed, meaning that the modulation arises from photons corresponding to the peak of the pulse profile \citep[see Section~\ref{sec:intro} and][]{an20, clark21, sim24}. 
We manually identified on- and off-pulse intervals,
%``by eye", 
selected only photons corresponding to these intervals, and then searched for orbital modulation as described.
Overall, searching for GOM with on-pulse photons proved to be more sensitive than just using all detected photons.
For example, for PSR J2215+5135 using all spin phases yields $H=11.97$, while selecting on-pulse photons results in a higher GOM significance $H=14.83$.

To ensure a robust and user-independent analysis for selecting on-pulse intervals,  we also adopted a reweighting procedure introduced by \cite{kerr19} and \cite{clark23}.
This method is based on the pulse shape and modifies the prior weights, $w$ (\texttt{MODEL\_WEIGHT}) into posterior weights $w'$,
\begin{equation*}
w'=\frac{wf(\phi)}{wf(\phi)+1-w},
\end{equation*}
where $f(\phi)$ is the pulsar's template pulse profile.
The pulse profile was modeled using a series of wrapped Gaussian functions, with parameters optimized by maximizing log-likelihood, as described in \cite{3PC}.
After folding and binning those reweighted photons to obtain orbital light curves, we refer to them as ``reweighted counts".
When the available timing solution does not allow for coherent spin-phase folding (e.g., in the case of PSR~J1227-4853, Sec.~\ref{sec:J1227}), we refer to the counts in the orbital light curves as ``weighted counts" (since they are still weighted according to the photon arrival directions and energies).

The 3PC catalog and data products that we used include 12 years of Fermi-LAT data,
from 2008 to 2020.
For the systems with GOM significance $>2.5\sigma$, we attempted to assess the significance of the GOM using all the available data. 
The primary obstacle is the validity interval of the publicly available timing solutions.
Many spiders have long-term orbital period variations \citep[see, e.g., ][]{pletsch15, shaifullah16, clark21}, which complicate expanding the ephemerides beyond their validity intervals. 

We chose to extrapolate data only for systems with stable timing solutions, i.e., those without high-order ($>2$) orbital period derivatives or \texttt{ORBIFUNC} terms.
The stability of pulsations was assessed by verifying the presence of pulsations beyond the formal validity range of the timing solution.
For these systems (PSR J1124-3653, PSR J1946-5403, and PSR J2256-1024), we analyzed Fermi-LAT data collected between 2008 August 4 and 2025 February 14.
Using the same photon selection as described above,  \texttt{P8R3\_SOURCE\_V3} \citep{atwood13, bruel18} instrument response, and by generating a 4FGL-DR4-based \citep{ballet23} source model for the RoI with the \texttt{LATSourceModel/make4FGLxml}\footnote{\url{https://github.com/physicsranger/make4FGLxml}} package, we calculated prior photon weights using \texttt{gtsrcprob}.
These weights were then reweighted as described previously.

Finally, to compare the MFs with gamma-ray fluxes in Section \ref{sec:discussion}, we calculated the $0.1-300$ GeV and $0.1-1$ GeV gamma-ray fluxes of the transitional systems PSR J1023+0038 and PSR J1227-4853, using separately the time intervals before and after their state transitions (Table \ref{tab:properties}).
To do so, we performed a standard binned-likelihood analysis using \texttt{fermipy} \citep{wood17}.
We selected \texttt{SOURCE} class photons within a $15^\circ$ RoI, with energies between $0.1$ and $300$ GeV and zenith angles $<90^\circ$, and used \texttt{P8R3\_SOURCE\_V3} instrument response files.
The sky model was based on 4FGL catalog sources, together with the Galactic diffuse model (\texttt{gll\_iem\_v07.fits}) and isotropic background (\texttt{iso\_P8R3\_SOURCE\_V3\_v1.txt}) models. 
The gamma-ray spectra of PSR J1023+0038 and PSR J1227-4853 were fitted with a \texttt{PLSuperExpCutoff2} spectral model.
During the fit, we allowed the normalizations of sources within $5^\circ$, the diffuse background normalization, and the spectral parameters of PSR J1023+0038 and PSR J1227-4853 to vary.

\section{Results}
\label{sec:results}

\begin{figure}[ht!]
\includegraphics[width=0.5\textwidth]{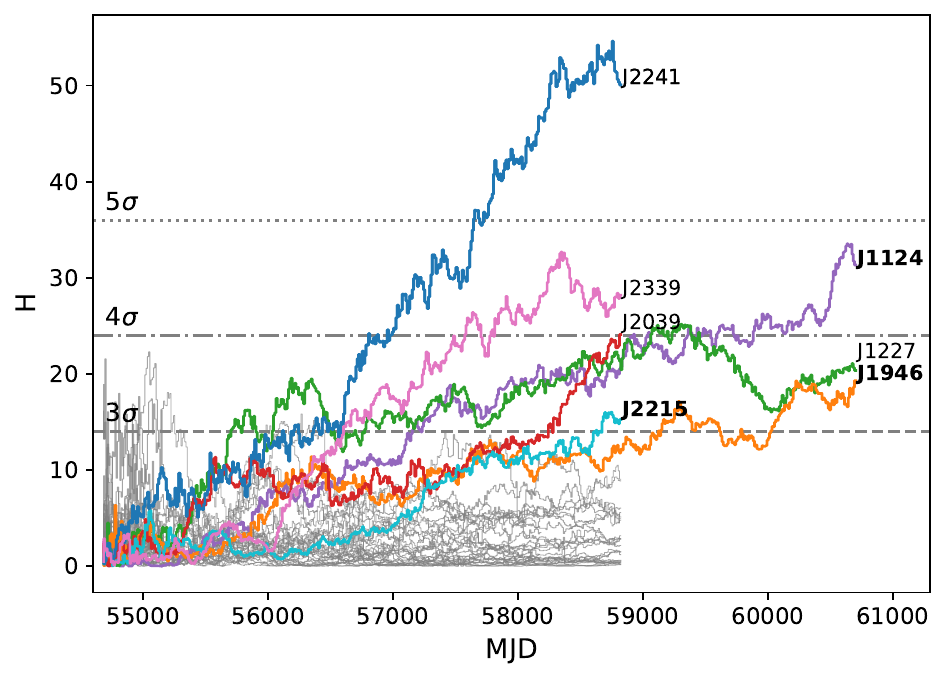 }
\caption{Weighted H-test significance for the seven systems with detected GOM, accumulated over time, using reweighted $0.1-1$ GeV photons (except for PSR J1227-4853, where regular photon weights were used). Gray lines indicate systems where no significant GOM was detected. The dashed horizontal lines represent the thresholds for a 3$\sigma$, 4$\sigma$, and 5$\sigma$ confidence levels. Systems with newly discovered GOM are shown in bold.
\label{fig:Htest}}
\end{figure}

We have detected significant soft pulsed GOM ($\geq3\sigma$; $0.1-1$ GeV band) in seven spider pulsars: four redbacks and three black widows (see Table \ref{tab:results}).
We show in Figure \ref{fig:lcs} the orbital gamma-ray light curves of these systems (in the $0.1-1$ GeV energy band, and using reweighted photons as explained in Section~\ref{sec:LAT}).
The remaining 36 folded light curves, where we did not detect GOM, can be found in Appendix~\ref{sec:appendixLCs} (Figure \ref{fig:all_lcs}).
The time evolution of the H statistic for these seven systems is shown in Figure \ref{fig:Htest}.
As Fermi-LAT collects more data, the significance of GOM detection increases over time.

To verify that the observed modulation is not caused by exposure variations, we folded the good-time intervals at the orbital period.
The relative variations in exposure time with orbital phase do not exceed $\sim0.5\%$, which is too small to account for the observed modulations.

We also investigated whether the GOM has an unpulsed component.
For each of the seven systems with detected GOM, we selected off-pulse intervals based on the pulse profiles and repeated the search for orbital modulation using only off-pulse photons.
In all cases, no significant modulation was detected in the off-pulse data.
This indicates that the observed GOM is pulsed.

Thus, we have discovered GOM in three new spider pulsars: PSR~J2215+5135 (redback), PSR~J1124-3653  and PSR~J1946-5403 (both black widows).
This nearly doubles the population of spiders with detected soft pulsed GOM.
The detected systems show MFs in the range of 18-43\% (Figure~\ref{fig:lcs} and Table \ref{tab:results}).
For the spiders without significantly detected GOM, we place upper limits on the MF (in the range 16-100\%, depending mostly on source brightness, see below).

In the following, we present our results on all three systems in which we have discovered GOM (Sec.~\ref{sec:newdet}), seven systems with previously reported GOM (Sec.~\ref{sec:prevdet})
and five eclipsing spiders (for which we provide MF upper limits, Sec.~\ref{sec:eclipsing}).

\begin{deluxetable*}{lcccc}
\tablewidth{\textwidth}
\tablecaption{Significance of GOM, modulated fraction (or upper limits) and phase of maximum flux for the pulsars studied, using reweighted $0.1-1$ GeV photons. The time range used for the analysis is shown in the last column. Systems where we discover GOM are highlighted in bold.
%and analyzed data time range. 
\label{tab:results}}
% \tablehead{
% \multicolumn{1}{l}{PSR} &
% \colhead{Significance, $\sigma$} & 
% \colhead{Modulated Fraction, \%} & 
% \colhead{$\Phi_{max}$} & 
% \colhead{Start Finish, MJD} \\
% & & & &
% \colhead{MJD}
% }
\tablehead{
PSR & Significance, $\sigma$ & Modulated Fraction, \% & $\Phi_{max}$ & Start - Finish, MJD
}
\startdata
\textbf{J1124-3653} * & 4.6 & 33.7$^{+9.0}_{-9.1}$ & $0.245\pm0.040$ & 54682-60720\\
J1227-4853 (t) \textdagger * [1]& 3.6 & $18.0^{+5.4}_{-5.3}$ & $0.133\pm0.040$ & 54682-60720\\
% J1946-5403 & 2.8 & $27.4\pm11.2$ & $0.174\pm0.061$ \\
\textbf{J1946-5403} * & 3.4 & $28.7\pm10.0$ & $0.133\pm0.051$ & 54682-60720\\
J2039-5617 [2] & 4.0 & $42.5^{+12.6}_{-12.9}$ & $0.285^{+0.051}_{-0.040}$ & 54682-58839\\
\textbf{J2215+5135} & 3.0 & $20.6\pm8.0$  & $0.255\pm0.061$ & 54682-58839\\
J2241-5236 [3] & 6.1 & $18.3\pm5.2$ & $0.295\pm0.040$ & 54682-58839\\
% J2256-1024 & 2.6 & $34.9^{+14.9}_{-15.2}$  & $0.386^{+0.071}_{-0.061}$ \\
% J2256-1024* & 1.0 & $<$37.3 & - \\
J2339-0533 [4] & 4.4 & $20.7^{+5.8}_{-5.9}$  & $0.295\pm0.040$ & 54682-58839\\
\hline
J0023+0923 & 0.2 & $<$52.5 & - & 54682-58839\\
J0251+2606 & 0.3 & $<$66.6 & - & 54682-58839\\
J0312-0921 & 1.3 & $<$51.6 & - & 54682-59224\\
J0610-2100 & 1.8 & $<$33.1 & - & 54682-58839\\
J0636+5128 & 0.1 & $<$80.0 & - & 54682-58836\\
J0952-0607 & 1.5 & $<$100 & - & 54682-58839\\
J1023+0038 (t) & 0.1 & $<$22.1 & - & 54682-56500\\
J1048+2339 (e) & 0.5 & $<$100 & - & 56509-57549\\
J1221-0633 & 1.5 & $<$66.7 & - & 54682-58839\\
J1301+0833 & 1.0 & $<$41.9 & - & 54682-58839\\
J1302-3258 & 0.1 & $<$43.9 & - & 54682-58839\\
J1431-4715 & 0.8 & $<$89.5 & - & 54682-58839\\
J1446-4701 & 1.0 & $<$80.1 & - & 54682-58839\\
J1513-2550 & 0.2 & $<$44.2 & - & 54682-58839\\
J1544+4937 & 1.1 & $<$100 & - & 54682-58839\\
J1555-2908 (e) & 0.2 & $<$49.2 & - & 54682-58493\\
J1622-0315 & 0.2 & $<$100 & - & 54682-58839\\
J1627+3219 & 0.5 & $<$71.4 & - & 54682-58839\\
J1628-3205 & 0.7 & $<$60.5 & - & 54682-57300\\
J1641+8049 & 0.5 & $<$90.4 & - & 54682-58190\\
J1653-0158 & 0.2 & $<$16.1 & - & 54682-58839\\
J1803-6707 & 0.7 & $<$78.0 & - & 54682-58839\\
J1805+0615 & 1.5 & $<$78.6 & - & 54682-58839\\
J1810+1744 & 2.1 & $<$15.7 & - & 54682-58839\\
J1816+4510 (e) & 0.6 & $<$32.0 & - & 54682-58839\\
J1833-3840 & 1.2 & $<$100 & - & 57313-58193\\
J1908+2105 & 1.0 & $<$84.1 & - & 54682-58693\\
B1957+20 (e) & 0.5 & $<$30.1 & - & 54682-57976\\
J2017-1614 & 0.2 & $<$60.2 & - & 54682-58839\\
J2047+1053 & 0.6 & $<$100 & - & 54682-58839\\
J2051-0921 & 1.3 & $<$100 & - & 54682-57874\\
J2052+1219 & 1.7 & $<$100 & - & 54682-58839\\
J2129-0429 (e) & 2.2 & $<$63.9 & - & 54682-58839\\
J2214+3000 & 1.0 & $<$18.9 & - & 54682-58839\\
J2234+0944 & 0.1 & $<$47.8 & - & 54682-58839\\
J2256-1024 & 1.1 & $<$36.7 & - & 54684-60720\\
\hline
\enddata
\tablecomments{Different source types and analysis are indicated: (t) - transitional MSPs, (e) - eclipsing, * - extended analysis beyond 3PC timing solution, \textdagger - J1227 using prior photon weights and timing solution from \cite{demartino20} (see Sections~\ref{sec:LAT} and \ref{sec:results} for details). Discovery references: [1] - \cite{an22}, [2] - \cite{ng18, clark21}, [3] - \cite{an18}, [4] - \cite{an20}}
% \tablenotetext{*}{text}
\end{deluxetable*}

\subsection{Three new discoveries}
\label{sec:newdet}

\subsubsection{PSR J1124-3653}
\label{sec:J1124}

PSR J1124-3653 (J1124) is a black widow pulsar with $P_\textrm{s}=2.41$ ms  in a $5.45$ hr orbit.
In X-rays, J1124 exhibits orbital variability, with a maximum at the pulsar's inferior conjunction \citep{gentile14, sim2024}.

From our gamma-ray study, we have discovered a sinusoidal modulation from J1124, with a significance of $3.9\sigma$ ($H=22.7$).
The gamma-ray pulsations of J1124 are detectable far beyond the 3PC timing solution interval, as we show in Appendix~\ref{sec:exteded_timing} (Figure \ref{fig:J1124_J1946_timing}).
Given the stability of the timing solution, we extended the search beyond the original validity range of the ephemeris to include the full 2008-2025 time range.
Using this extended search, in the $0.1-1$ GeV band with reweighted photons, we find a GOM significance of $4.6\sigma$ ($H=31.42$).
The maximum of the GOM coincides with the pulsar's superior conjunction ($\Phi_{max}=0.24\pm0.04$) and the MF is $34\pm9\%$.

\subsubsection{PSR J1946-5403}
\label{sec:J1946}

PSR J1946-5403 (J1946) is a black widow pulsar with $P_\textrm{s}=2.71$ ms and $P_\textrm{b}=3.12$ hr \citep{camilo15}.
It shows XOM with two peaks centered around pulsar's superior conjunction \citep{sim2024}. 
This suggests that the companion wind is weaker than that of the pulsar, and the IBS is wrapped around the companion star.

From our standard study in the $0.1-1$ GeV band using reweighted photons within the 3PC timing solution, we found indications of GOM, with a significance of $2.8\sigma$.
% Given the stability of its orbital period, we extended data beyond the original ephemeris validity range.
By extending the analysis beyond the time range of the 3PC ephemeris(from August 2008 to March 2018), we continued detecting pulsations, indicating that the timing solution remains stable over a longer timescale (Figure \ref{fig:J1124_J1946_timing}).
Including these additional data, we detect GOM with a significance of  $3.1\sigma$ ($H=15.51$).
We thereby measure a MF of $26\pm10\%$ and a peak phase of $\Phi_{max}=0.13\pm0.06$.

\subsubsection{PSR J2215+5135}
\label{sec:J2215}
PSR J2215+5135 (J2215) is a redback system with a spin period $P_\textrm{s}=2.61$ ms and $P_\textrm{b}=4.14$ hr. 
From an optical study of its strongly irradiated companion,  \cite{linares18} reported one of the most massive known neutron stars with $M_{NS}\simeq2.3M_{\odot}$.
This system also exhibits XOM \citep{sullivan25},  characterized by two peaks near the pulsar's inferior conjunction, $\Phi=0.75$. 

We found soft GOM from J2215 with a significance of $3.0\sigma$ ($H=15.12$) using reweighted photons in the $0.1-1$ GeV energy range.
The MF of J2215 is $21\pm8\%$ and the peak is at the pulsar's superior conjunction $\Phi_{max}=0.26\pm0.06$.
Due to stochastic orbital period variations, it was not possible to expand the dataset beyond the analyzed 3PC interval.

\subsection{Previously detected systems}
\label{sec:prevdet}

Significant soft GOM had previously been reported in four spider systems: three redbacks (PSR J1227-4853, \citealt{an22}; PSR J2039-5617, \citealt{clark21}; PSR J2339-0533, \citealt{an20}) and one black widow PSR J2241-5236 \citep{an18}.
We confirm all four detections and present the results of our reanalysis in this section.

\subsubsection{PSR J1227-4853}
\label{sec:J1227}

PSR J1227-4853 (J1227) is a transitional MSP with $P_\textrm{s}=1.69$ ms $P_\textrm{b}=6.91$ hr.
At the end of 2012, it transitioned from a Low-mass X-ray Binary (LMXB) to a pulsar state \citep{bassa14}.
During the pulsar state, XOM with maximum at the neutron star's inferior conjunction was observed with XMM-Newton and NuSTAR \citep{demartino15, demartino20}.

Using timing solutions for the orbit of J1227, including time of ascending node, binary period, and its derivatives, \cite{an22} found GOM in the $60-1000$ MeV energy band using 12.5 years of Fermi-LAT data, spanning from 2008 August to 2021 February. 
Interestingly, the modulation was detected in both the LMXB and the MSP states. 

The 3PC timing solution is valid only from 2014 February to 2017 August (MJD 56707-57970), and we detect gamma-ray pulsations %are detected 
only within this time range. 
As a result, reweighting can be applied only within this specific validity interval of the timing solution.
Within this time range, we do not detect significant GOM using reweighted photons in the $60-1000$ MeV, $0.1-1$ GeV, or $0.1-300$ GeV energy bands. 
Similarly, using prior probability weights, we do not detect significant modulation in any of these energy bands.

The 3PC timing solution contains 11 orbital frequency derivatives, which are necessary to precisely phase-align pulsations while accounting for stochastic orbital period variation. 
This prevents reliable extrapolation of the orbital phase model \citep{clark21}.

However, using the radio timing solutions for the orbit from \citet[which does not incorporate information about the pulsar spin]{an22} and prior photon weights, along with the time range from 2008 August to 2025 February, we confirm the detection of significant GOM in the $60-1000$ MeV and $0.1-1$ GeV energy bands.
In the latter energy range (our standard band), we measure $H=19.92$ ($\sim3.6\sigma$), MF$=18\pm5\%$ and $\Phi_{max}=0.13\pm0.04$. 

\begin{figure}[ht!]
\includegraphics[width=0.5\textwidth]{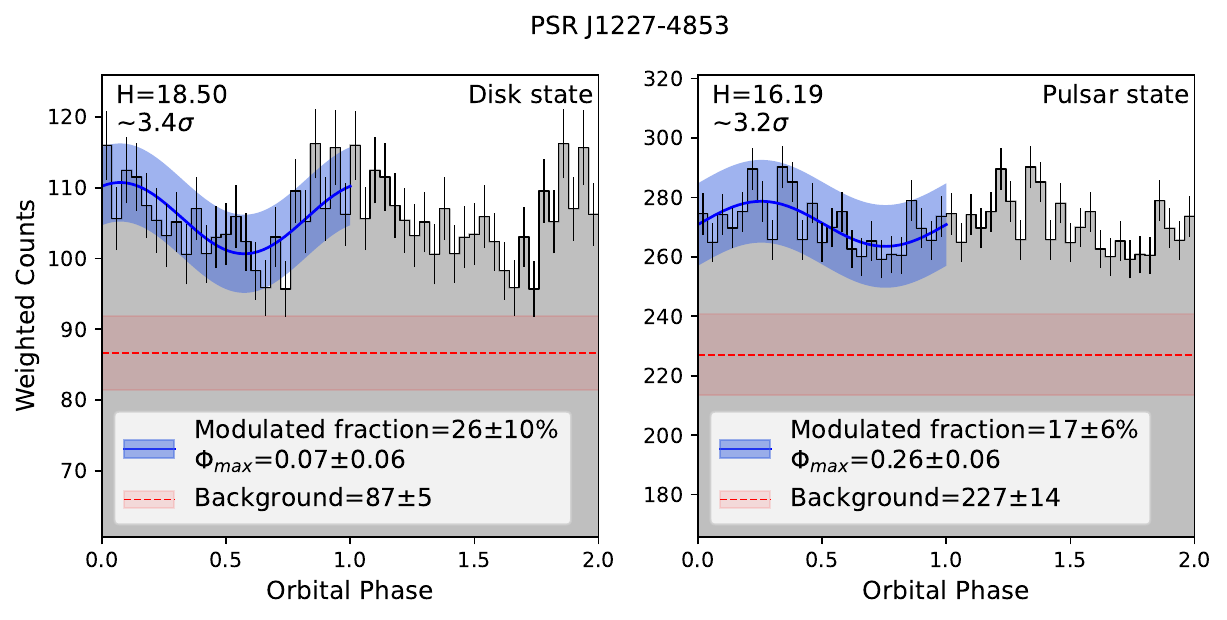}
\caption{Weighted, phase-folded gamma-ray light curves of J1227 in the $0.1-1$ GeV band, in the disk state (left) and the pulsar state (right).
\label{fig:J1227_disk_vs_psr}}
\end{figure}

To study the modulation in the two different states of J1227, we split the LAT dataset at the time of the state transition \citep[MJD $56250$,][]{bassa14}. 
The data before this date correspond to the disk state, while the data after correspond to the pulsar state.
We detect significant GOM  in both states, consistent with the results reported by \cite{an22}, with $H=18.50$ ($\sim3.4\sigma$) in the disk state and $H=16.19$ ($\sim3.2\sigma$) in the pulsar state (Figure \ref{fig:J1227_disk_vs_psr}).
Despite the state transition being accompanied by a drop in flux, the MFs remain comparable:  $MF=26\pm10\%$ before and $MF=17\pm6\%$ after the transition (see Section~\ref{sec:discussion_transitionals}). 
However, while the maximum in the pulsar state is located at the pulsar's superior conjunction ($\Phi_{max}=0.26\pm0.06$), we find that the peak in the disk state occurs earlier, at $\Phi_{max}=0.07\pm0.06$ (as can be seen in Figure~\ref{fig:J1227_disk_vs_psr}).

\subsubsection{PSR J2039-5617}
PSR J2039-0533 (J2039) was the first redback MSP system discovered through a blind search for gamma-ray pulsations \citep{clark21}.
It has $P_\textrm{s}=2.65$ ms and $P_\textrm{b}=5.5$ hr. 
J2039 exhibits a typical double-peaked X-ray orbital light curve, centered around the pulsar's inferior conjunction \citep{salvetti15, romani15}.
From optical light curve modeling, \cite{clark21} estimated an orbital inclination of $\sim 75^\circ$.

Before the discovery of pulsations, \cite{ng18} found sinusoidal orbital modulation in the gamma-ray flux of 3FGL J2039.6–5618, peaking at the pulsar's superior conjunction—half an orbit away from the X-ray maximum.
\cite{clark21} confirmed this detection, showing that the orbitally modulated signal is pulsed. 
They produced an orbital light curve for photons with energies higher than $100$ MeV without selecting or reweighting photons based on the pulse profile or phase, reporting a MF of $24\pm5\%$ and a peak phase $\Phi_{max}=0.25\pm0.03$.

In the $0.1-300$ GeV energy band, we also detect GOM with a significance of $4.0\sigma$, a MF of $25\pm8\%$ and $\Phi_{max}=0.26\pm0.05$, in agreement with the results of \cite{clark21}.
Using reweighted $0.1-1$ GeV photons, we measure a higher MF of $43\pm13\%$ with $\Phi_{max}=0.29\pm0.05$ with a significance of $4.0\sigma$.
This suggests that the modulation is stronger at lower energies ($0.1-1$ GeV), while higher energy photons ($>1$~GeV) introduce additional flux that is not modulated.

\subsubsection{PSR J2241-5236}

PSR J2241-5236 (J2241) is a black widow pulsar with $P_\textrm{s}=2.2$ ms and $P_\textrm{b}=3.5$~hr. 
\cite{an18} found low-energy ($<1$ GeV) orbital modulation in J2241, with the first, stronger peak occurring near the pulsar's superior conjunction and a second, smaller peak near the pulsar's inferior conjunction. 
To explain this, \cite{an18} suggested that the strength ratio of the pulsar and companion winds varies, so that the IBS wraps around the companion, but occasionally also around the pulsar. 
This would produce two peaks at the pulsar's superior and inferior conjunctions.

A 20 ks Chandra observation of J2241 revealed thermal X-ray emission, presumably originating from hot polar caps on the neutron star's surface, and a possible nonthermal component \citep{keith11, an18}.
However, the observation was not sensitive enough to probe XOM \citep{an18}.
% No significant non-thermal component was detected in the Chandra data \citep{an18}.
Using NICER, \cite{guillot19} detected thermal X-ray pulsations from J2241.

We also detect GOM with a significance of $6.1\sigma$ using reweighted $0.1-1$ GeV photons. 
Since the orbital light curve of this system is peculiar, our sinusoidal model does not fit the data well, as expected (see Figure \ref{fig:lcs}).
However, it provides reasonable estimates for the MF ($18\pm5\%$) and the location of the stronger peak ($\Phi_{max}=0.30\pm0.04$).

\begin{figure*}[ht!]
\centering
\includegraphics[width=0.85\textwidth]{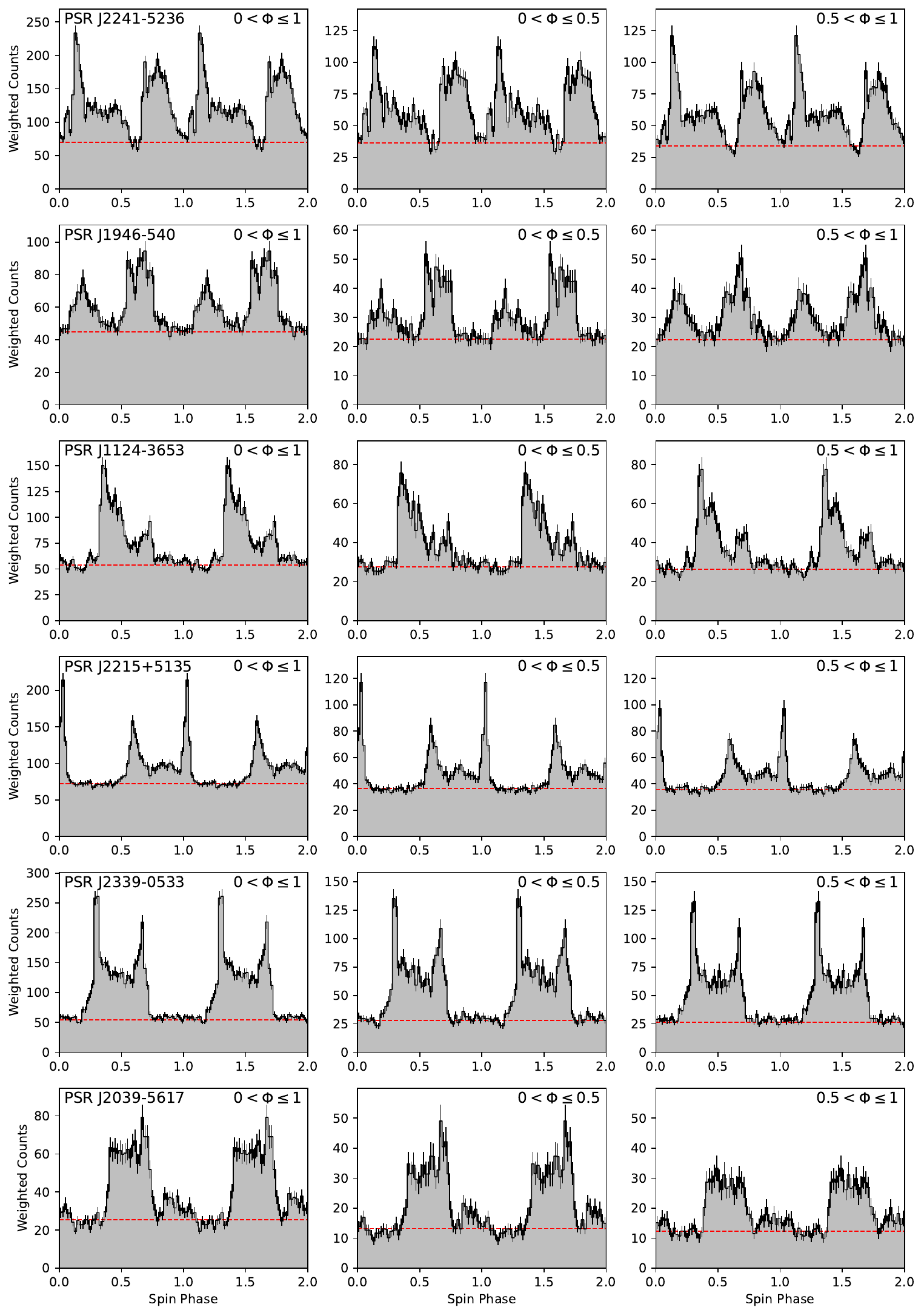 }
\caption{Pulse profiles of six pulsars with detected GOM in the $0.1-300$ GeV band, measured in three orbital phase selections: orbit-averaged ($0<\Phi\leq 1$), near pulsars superior conjunction ($0<\Phi\leq 0.5$), and at pulsars inferior conjunction ($0.5<\Phi\leq 1$).
\label{fig:pulse_profiles}}
\end{figure*}

\subsubsection{PSR J2339-0533}
The redback pulsar PSR J2339-0533 (J2339) is a $2.9$ ms pulsar in a $4.6$ hr orbit. 
The X-ray light curve of J2339 is double-peaked around the pulsar's inferior conjunction \citep{romani11, kandel19}.

\cite{an20} discovered sinusoidal GOM of J2339 in the $100-600$ MeV energy band by selecting pulsed photons. 
The gamma-ray orbital light curve is anticorrelated with the X-ray light curve, similar to J2039. 
The detection has a false-alarm probability of $p\approx10^{-7}$.
The authors also note that the significance is relatively sensitive to the energy selection, reporting $p=2\times10^{-4}$ for the $0.1-1$ GeV energy band.

Using the same selection of spin phase as \cite{an20} for $0.1-1$ GeV photons, we also detect GOM with $H=19.97$, corresponding to $p=3\times10^{-4}$, consistent with \cite{an20}.
With our standard analysis using reweighted $0.1-1$ GeV photons, we find $H=28.23$ ($\sim4.4\sigma$) with $\Phi_{max}=0.30\pm0.04$ and a MF of $21\pm6\%$.

\subsection{Eclipsing binaries}
\label{sec:eclipsing}

To maximize sensitivity for detecting gamma-ray eclipses, we used reweighted photons in the full Fermi-LAT energy band of $0.1-300$ GeV as explained by \cite{clark23}, that provide sufficient photon statistics.
Eclipses were confirmed in PSRs B1957+20, J1048+2339, J1555-2908, J1816+4510, and J2129-0429, consistent with the results of \citet[Figure \ref{fig:eclipsing}]{clark23}.

After confirming the presence of eclipses, we used our standard analysis in the soft band ($0.1-1$ GeV) to search for GOM. We did not detect significant GOM from any of the five eclipsing systems, and we placed upper limits on their MFs. 
The folded light curves in the soft band are shown in Figure \ref{fig:eclipsing_soft}.

\subsubsection{PSR J1048+2339}

The redback MSP PSR J1048+2339 (J1048) has $P_\textrm{s}=4.67$ ms and $P_\textrm{b}=6.01$ hr.
Its 3PC timing solution spans $\sim3$ yr and exhibits significant orbital period variations, making it difficult to extend over longer timescales.
However, it remains sufficient to detect gamma-ray pulsations and observe an eclipse of the pulsar by its companion \citep{clark23}.

We do not detect significant GOM from J1048.
Due to the short duration of the ephemeris and the limited number of detected photons, we are unable to place strong constraints on the MF, with an upper limit of $<100\%$.

\subsubsection{PSR J1555-2908}

PSR J1555-2908 (J1555) is one of the most energetic MSPs: a black widow pulsar with a spin-down luminosity $\dot{E}>3.1\times10^{35}$ erg s$^{-1}$ and $P_\textrm{s}=1.79$ ms, $P_\textrm{b}=5.6$ hr. 

The timing solution of J1555 spans from 2008 August to 2019 January, and we detect significant pulsations throughout this whole time range.
However, no significant GOM is detected, with an upper limit on the MF of $49\%$.

\subsubsection{PSR J1816+4510}

PSR J1816+4510 (J1816) is a redback pulsar with $P_\textrm{b}=8.66$ h and $P_\textrm{s}=3.19$ ms.
J1816 has one of the brightest and hottest companions among spiders with an effective temperature of $\gtrsim15000$ K \citep{kaplan12}.
It shows both radio and gamma-ray eclipses \citep{stovall14, clark23}.

We do not detect GOM from this system and we place a $3\sigma$ upper limit on the MF of $<32\%$.

\subsubsection{PSR B1957+20}

PSR B1957+20 (B1957) is the first discovered black widow system: a MSP with a $P_\textrm{s}=1.61$ ms pulsar in a $9.17$ hr orbit around an ultra-light companion. 
Chandra observations revealed a double-peaked X-ray orbital light curve with maximum emission occurring around the pulsar's superior conjunction \citep{huang12}, unlike most (redback) spiders \citep{wadiasingh17}.

Gamma-ray analysis with Fermi-LAT provided evidence for orbital-phase-dependent emission from B1957. 
\cite{wu12} analyzed three years of Fermi-LAT data and reported spectral variations across the orbit, identifying an additional gamma-ray component above $\sim2.7$ GeV that appeared at the pulsar inferior conjunction.
While their analysis suggested potential orbital modulation at low energies, the significance was only $\sim2.3\sigma$.
\cite{corbet22} did not find any significant modulation in this system.

In our study, we analyzed Fermi-LAT data from the $\sim11$ yr period where the 3PC timing solution was valid. We do not find significant GOM from B1957.
Using reweighted $0.1-1$ GeV photons, we place a $3\sigma$ upper limit of $30\%$ on the MF.

\subsubsection{PSR J2129-0429}
The redback system PSR J2129-0429 (J2129) has $P_\textrm{s}=7.61$ ms and $P_\textrm{b}=15.25$ hr.
The X-ray orbital light curve obtained with XMM-Newton and NuSTAR revealed a double-peaked maximum around the pulsar's inferior conjunction \citep{hui15, kong18}. 

For PSR J2129-0429 (J2129), the full energy band light curve ($0.1-300$~GeV) shows the eclipse at phase 0.25 \citep{clark23}.
Our analysis in the soft energy band yields tentative evidence of sinusoidal orbital modulation (Figure \ref{fig:all_lcs}).
However, this modulation is not statistically significant ($2.2\sigma$ in the $0.1-1$ GeV energy band when using reweighted photons), so we place an upper limit of $<64\%$ on the MF.

\begin{deluxetable*}{lcccc}[ht!]
\tablewidth{\textwidth}
\tabletypesize{\small}
\tablecaption{Spectral parameters for the seven spiders with detected GOM (Sec.~\ref{sec:spec} for details)
%from a \texttt{PLSuperExpCutoff2} model fit: power law index $\gamma_1$, exponential factor $a$) 
and $0.1-300$ GeV gamma-ray flux around orbital flux maximum and minimum.
\label{tab:seds}}
\tablehead{
Name & Orbital Phase & $\gamma_1$ & $a/10^{-3}$ & $F_\textrm{g}/10^{-12}$ erg cm$^{-2}$ s$^{-1}$
}
\startdata
J1124-3653 & $0<\Phi\leq0.5$ & $-1.57\pm0.05$ & $5.1$ & $15.3\pm0.7$ \\
 & $0.5<\Phi\leq1$ & $-1.45\pm0.06$ & $5.1$ & $11.8\pm0.7$ \\
\hline
J1227-4853 & $0.75<\Phi\leq1.25$ & $-2.22\pm0.07$ & $3.1$ & $37.1\pm1.8$ \\
(disk state) & $0.25<\Phi\leq0.75$ & $-2.02\pm0.07$ & $3.1$ & $31.1\pm1.7$ \\
\hline
J1227-4853 & $0<\Phi\leq0.5$ & $-1.88\pm0.06$ & $4.7$ & $16.5\pm0.9$ \\
(pulsar state) & $0.5<\Phi\leq1$ & $-1.73\pm0.07$ & $4.7$ & $13.4\pm0.8$ \\
\hline
J1946-5403 & $0<\Phi\leq0.5$ & $-1.51\pm0.07$ & $8.3$ & $11.7\pm0.6$ \\
 & $0.5<\Phi\leq1$ & $-1.28\pm0.07$ & $8.3$ & $10.5\pm0.7$ \\
\hline
J2039-5617 & $0<\Phi\leq0.5$ & $-1.20\pm0.06$ & $9.0$& $16.3\pm0.8$ \\
 & $0.5<\Phi\leq1$ & $-0.95\pm0.08$ & $9.0$ & $11.9\pm0.8$ \\
\hline
J2215+5135 & $0<\Phi\leq0.5$ & $-1.60\pm0.06$ & $5.9$ & $18.5\pm0.8$ \\
 & $0.5<\Phi\leq1$ & $-1.50\pm0.06$ & $5.9$ & $16.2\pm0.8$ \\
\hline
J2241-5236 & $0<\Phi\leq0.5$ & $-0.75\pm0.04$ & $12.5$ & $26.7\pm0.9$ \\
 & $0.5<\Phi\leq1$ & $-0.60\pm0.05$ & $12.5$ & $22.8\pm0.8$ \\
\hline
J2339-0533 & $0<\Phi\leq0.5$ & $-1.13\pm0.04$ & $7.7$ & $30.3\pm1.0$ \\
 & $0.5<\Phi\leq1$ & $-1.0\pm0.03$ & $7.7$ & $28.0\pm0.9$ \\
\enddata
% \tablenotetext{*}{text}
\end{deluxetable*}

\subsection{Pulse profiles}

Previous studies have shown that the gamma-ray pulsar tends to be brighter around superior conjunction, corresponding to the orbital maximum of the GOM \citep{an20, clark21}.
To test this in our detected systems, we produced pulse profiles for six pulsars exhibiting GOM in three orbital phase intervals: orbit-averaged ($0<\Phi\leq1$), around the orbital maximum ($0<\Phi\leq0.5$), and around the orbital minimum ($0.5<\Phi\leq1$).
J1227 was not included because of its short ephemeris, during which gamma-ray pulsations are detected (see Section \ref{sec:J1227}). 
%
%It should be noted that 
The photon weights were assigned according to the orbit-averaged pulsar spectrum. 
The resulting pulse profiles are shown in Figure \ref{fig:pulse_profiles}.
We find that the profiles around superior and inferior conjunctions differ slightly in flux (weighted counts), but also in shape.
Further investigation via spin and orbital phase–resolved spectroscopy \citep[e.g.,][]{an20} or pulse profile modeling is beyond the scope of this work.

\begin{figure*}[!h]
\centering
\includegraphics[width=0.8\textwidth]{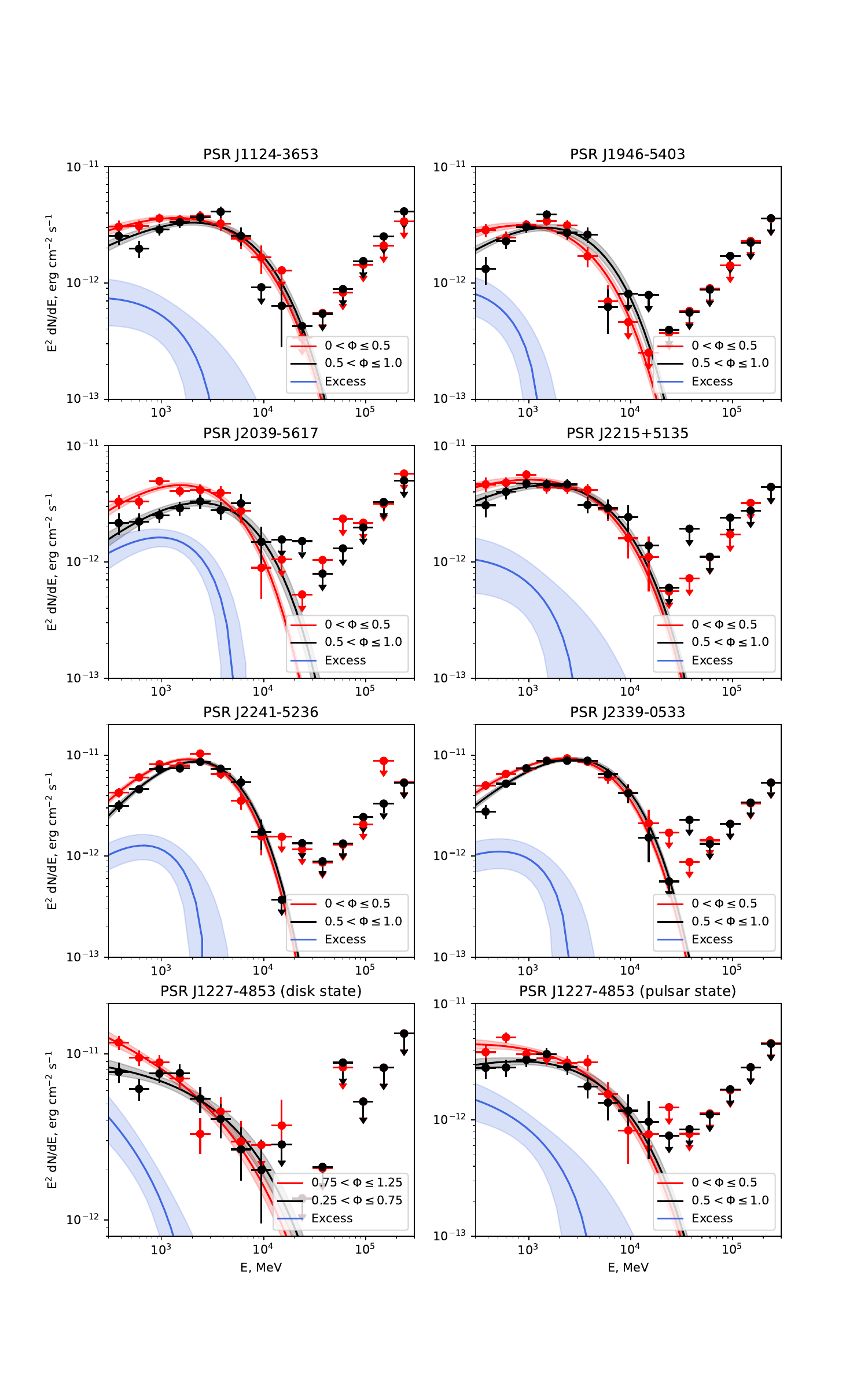}
\caption{Spectral energy distributions of pulsars with detected GOM, measured around their orbital flux maximum and minimum. Red and black points show the SEDs in these two phase intervals, respectively. Solid lines and shaded regions represent the best-fit models and their $1\sigma$ uncertainties. The blue curve shows the difference between the best-fit models for the two orbital phase selections.
\label{fig:seds}}
\end{figure*}

\subsection{Spectral variations}\label{sec:spec}
We also examined how the pulsar spectra change across the orbital phases in systems with detected GOM.
For each system, we performed a spectral analysis in the $0.3-300$ GeV  band  in two orbital phase intervals: $0 < \Phi \leq 0.5$ and $0.5 < \Phi \leq 1$ , corresponding to the pulsar’s superior and inferior conjunctions, respectively (shown as red and black data points in Figure \ref{fig:seds}) For the spectral analysis, we excluded data below $300$ MeV due to the large point-spread function and energy dispersion at low energies\footnote{\url{https://fermi.gsfc.nasa.gov/ssc/data/analysis/LAT_caveats.html}}.
The LAT spectra were modeled using a \texttt{PLSuperExpCutoff2} model\footnote{\url{https://fermi.gsfc.nasa.gov/ssc/data/analysis/scitools/source_models.html}},
$$\frac{dN}{dE}=N_0\left(\frac{E}{E_0}\right)^{\gamma_1}\exp(-aE^b),$$
with the pivot energy fixed at $E_0=1$ GeV and exponential index fixed at $b=2/3$ , as is typically done for gamma-ray pulsars \citep{abdollahi20, 3PC}.
The normalization $N_0$ and the power-law index $\gamma_1$ were left free to vary.
The exponential factor $a$ was fixed to its orbit-averaged value.
All other sources and diffuse backgrounds were not allowed to vary between the two selected phase intervals.

Since GOM of J1227 in the disk state does not peak near the superior conjunction (see Section \ref{sec:J1227}), for this system we selected orbital phase intervals $0.75<\Phi\leq1.25$ and $0.25<\Phi\leq0.75$.

The resulting spectral energy distributions (SEDs) are shown in Figure \ref{fig:seds} and the corresponding parameters are reported in Table \ref{tab:seds}.
The excess is calculated as the difference between the best-fit models at superior and inferior conjunction, and shown with blue lines in Figure~\ref{fig:seds} (with blue shaded $1\sigma$ error region).
For all systems the excess peaks at $\lesssim 1$ GeV and decays quite sharply above that energy, confirming that the detected GOM is soft.
In J1227, the excess is softer in the disk state than in the pulsar state.

\section{Discussion}
\label{sec:discussion}
We have detected significant GOM in seven out of the total of 43 spider systems analyzed.
We have discovered modulation in the pulsed gamma-ray flux of three spiders: one redback (J2215, Sec.~\ref{sec:J2215}) and two black widows (J1124 and J1946; Secs.~\ref{sec:J1124} and \ref{sec:J1946}, respectively).
In all three newly discovered cases, the GOM significance drops below $3\sigma$ when including the full LAT band ($0.1-300$ GeV), indicating that the modulation is soft.
This nearly doubles the number of spiders with detected soft and pulsed GOM (Table \ref{tab:results}).

Our results show that GOM in spiders is more common than previously thought. 
Since we conducted a systematic search across the spider population presented in 3PC, we estimated the significances accounting for trials, i.e., the search over the 43 independent systems.
For our new findings, PSR J1124-3653, J1946-5403 and J2215+5135, the post-trial significances are $3.8\sigma, 2.2\sigma$ and $1.6\sigma$, respectively.
The significance estimated using the H-test with one harmonic is higher (m=1, also known as the Rayleigh test; \citealt{buccheri83, deJager89}), yielding post-trial significances of $4.4\sigma, 2.6\sigma$ and $2.0\sigma$ for J1124, J1946 and J2215, respectively.
%PRE\&POST-TRIAL SIGNIFICANCE FROM H-test WITH ONE HARMONIC (m=1)?
%
Because the GOM shapes are clearly sinusoidal and their flux maxima align with those of all other spiders (near the pulsar's superior conjunction), we consider these detections to be robust.

\subsection{A universal modulated fraction}

\begin{figure}[h!]
\includegraphics[width=0.45\textwidth]{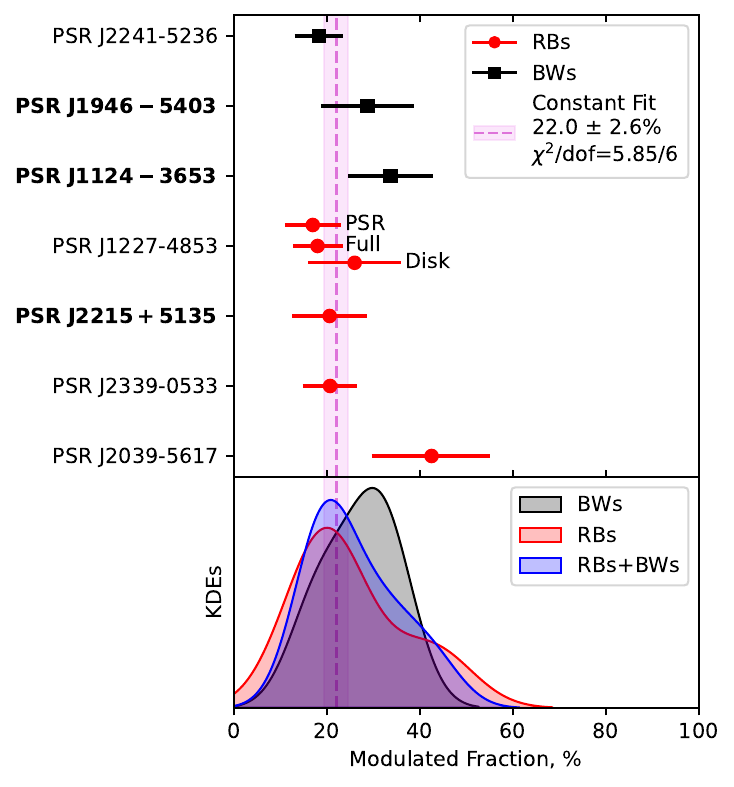}
\caption{{\it Top}: Distribution of modulated fractions in the 0.1-1 GeV energy band. {\it Bottom}: KDEs computed using Gaussian kernels %\texttt{scipy.stats.gaussian\_kde}, 
with bandwidth $7^{-1/5}$. %(determined by Scott’s Rule.
\label{fig:MFs}}
\end{figure}

\begin{figure}[h!]
\includegraphics[width=0.45\textwidth]{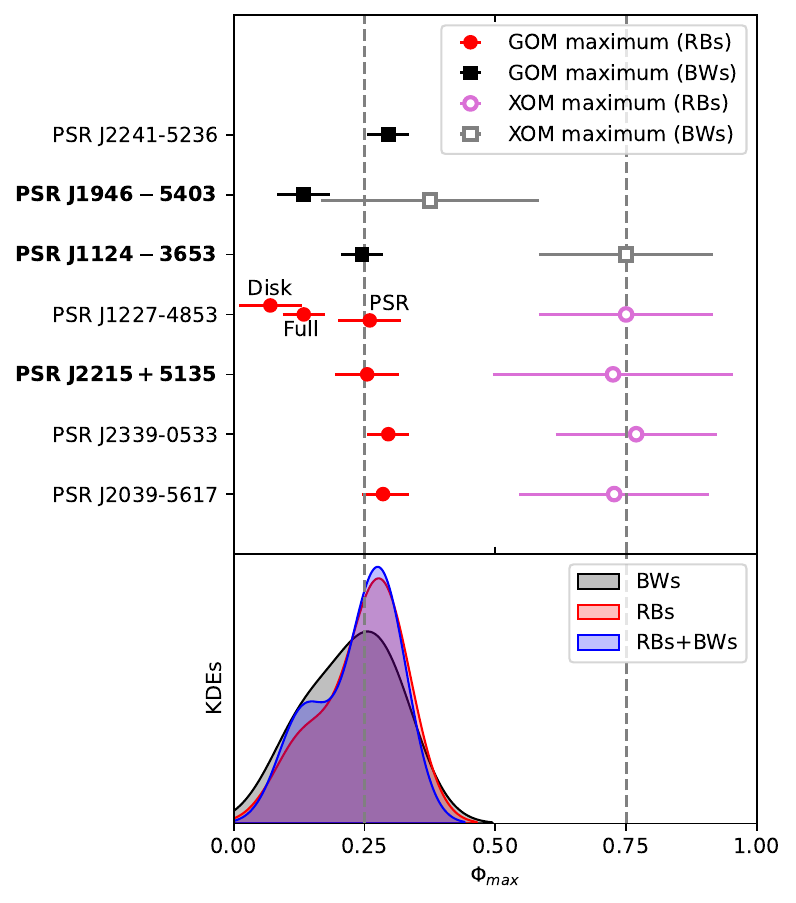}
\caption{{\it Top}: Distribution of gamma-ray and X-ray orbital modulation peak phases $\Phi_{max}$. {\it Bottom}: KDEs computed using Gaussian kernels %\texttt{scipy.stats.gaussian\_kde}, 
with bandwidth $7^{-1/5}$
\label{fig:PhiMax}}
\end{figure}

\begin{figure}[ht!]
\centering
\includegraphics[width=0.45\textwidth]{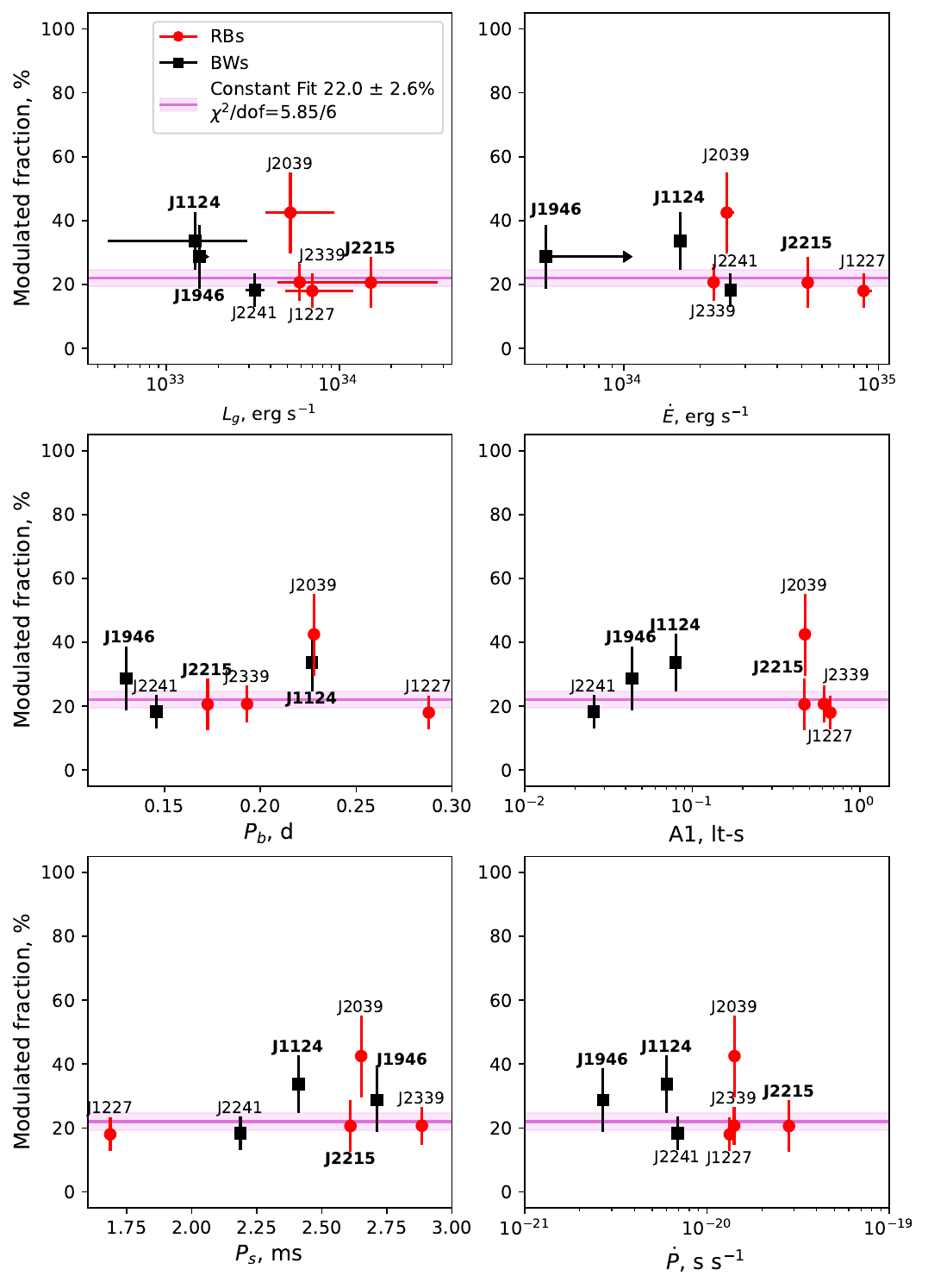}
\caption{Modulated fractions plotted versus $0.1-100$ GeV gamma-ray luminosity $L_g$, spin-down power $\dot{E}$, orbital period $P_\textrm{b}$, and projected semi-major axis of the pulsar $A1$, spin period $P_\textrm{s}$, and spin period derivative $\dot P$. 
\label{fig:MF_mulipanel}}
\end{figure}

We find that our measured MFs (Figure~\ref{fig:MFs}) and $\Phi_\textrm{max}$ (Figure~\ref{fig:PhiMax}) are relatively uniform across all systems, regardless of whether they are redbacks or black widows.
Figure \ref{fig:MFs} shows the distribution of MFs together with their kernel density estimates (KDEs).
Fitting the MFs with a constant yields a best-fit value of MF=$22.0\pm2.6\%$, with $\chi^2/\text{d.o.f.}=5.85/6$.
We also find that the MFs show no clear dependence on the spin period $P_\textrm{s}$, its derivative $\dot{P}$, orbital period $P_\textrm{b}$ (which differs by a factor $\simeq2$ across our sample) or other parameters such as gamma-ray luminosity $L_\textrm{g}$, spin-down power $\dot{E}$ or projected semi-major axis of the pulsar $A1$ (which differ by a factor $\gtrsim$10; see Figure \ref{fig:MF_mulipanel}).

The gamma-ray flux maxima of all seven detected systems are found near the pulsar's superior conjunction ($\Phi=0.25$ in our definition, see Figure \ref{fig:PhiMax}). 
XOM has previously been observed in all seven systems except for J2241.
Thus, we find that the redback J2215 and the black widow J1124 add to three other redbacks (J1227, J2039 and J2339) that exhibit an {\it anticorrelation} between  gamma-ray and X-ray fluxes (see Figure~\ref{fig:PhiMax} and references in Table~\ref{tab:properties}).
In these five systems, the GOM and XOM are {\it in antiphase}: the gamma-ray flux peaks near the pulsar's superior conjunction, while the X-ray flux peaks at inferior conjunction ($\Phi\simeq0.75$, Fig.~\ref{fig:PhiMax}).
In the black widow J1946, however, we find that GOM and XOM are {\it in phase}: both gamma-ray and X-ray modulations peak at the same orbital phase, near superior conjunction of the pulsar.


Redback companions are about ten times more massive and about twice as hot as black widow companions \citep[factor $\simeq 2$ higher base temperature,][]{turchetta23}.
The fact that the MFs are similar across all detected systems (regardless of whether they are redbacks or black widows) suggests a common underlying mechanism for the GOM, independent of the mass and temperature of the companion star.
MF and $\Phi_\textrm{max}$ seem also unrelated to the number of maxima in the optical orbital modulation (Table~\ref{tab:detected_gom}), suggesting that the gamma-ray modulation mechanism is independent of the irradiation state of the companion.
On the other hand, the phase of the X-ray flux maximum is thought to be determined by the IBS geometry or orientation \citep[Section~\ref{sec:intro};][]{wadiasingh17,wadiasingh18}.
Since we always find a soft GOM peak near superior conjunction, independently of the X-ray modulation maximum, we conclude that the origin of the GOM is not directly linked to the IBS orientation.

\begin{figure}[ht!]
\includegraphics[width=0.47\textwidth]{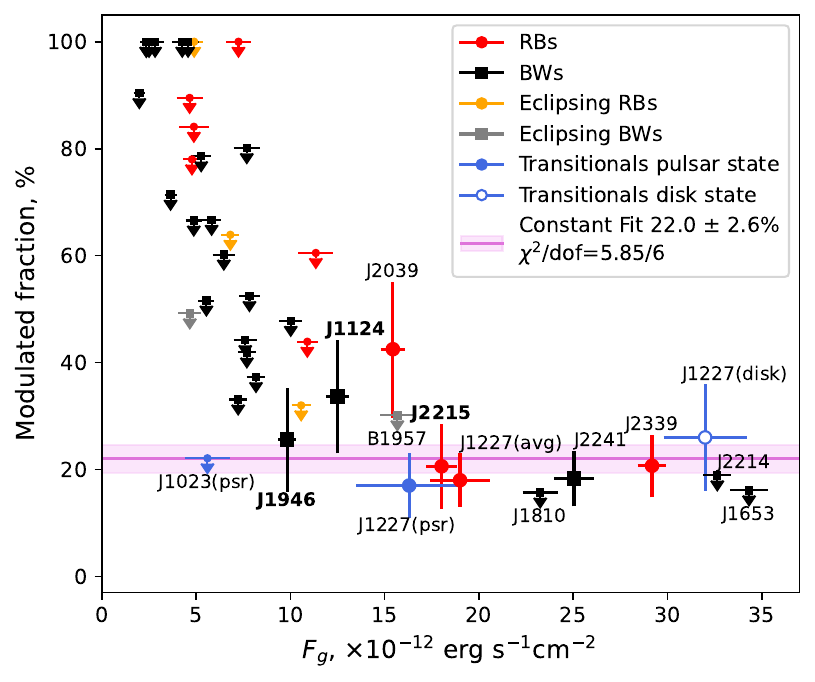}
\caption{Measured modulated fraction and upper limits as a function of $0.1-300$ GeV gamma-ray flux. 
\label{fig:MF_Fg}}
\end{figure}

Figure \ref{fig:MF_Fg} shows the measured MFs and upper limits, plotted against the gamma-ray flux of each system ($F_\textrm{g}$, in the 0.1-300~GeV band).
We show the MF upper limit and $F_\textrm{g}$ of J1023 in the pulsar state (MJD$<56500$), while for J1227 we plot the measured MFs and gamma-ray flux, both averaged and separately for disk and pulsar states.
Systems with $F_\textrm{g} > 1.5\times10^{-11}$~erg~s$^{-1}$~cm$^{-2}$ have GOM detections or  stringent upper limits on the modulated fraction (MF$\lesssim$30\%).
This is expected if the MF is independent of $F_\textrm{g}$, since brighter sources provide more photon counts, allowing for increased sensitivity to potential modulation.
Other factors that may contribute to the GOM detectability include the time spanned by the coherent timing solution
and the ``sharpness" of the pulse profile (Fig.~\ref{fig:pulse_profiles}).

Three black widows - PSR J1653-0158 (J1653), J1810+1744 (J1810) and J2214+3000 (J2214) - are not detected despite their high gamma-ray fluxes.
The corresponding $3\sigma$ upper limits on their MFs are $16.1\%$, $15.7\%$ and $18.9\%$ respectively.
These values are close to the constant fit level of $22.0\pm2.6\%$.
These upper limits are consistent with the average MF within $1-2~\sigma$, so we argue that the modulated fraction of the GOM in spider pulsars is universal.

\subsection{Transitional MSPs and GOM}
\label{sec:discussion_transitionals}
As shown by \cite{an22} and discussed in Section \ref{sec:J1227}, J1227 exhibits orbital modulation not only in the pulsar state, but also in the disk state.
Since gamma-ray pulsations from transitional MSPs have been detected in their pulsar state \citep{johnson15,3PC}, but not in the disk state \citep{3PC}, the additional gamma-ray emission in the disk state is presumably unpulsed.
However, the orbital modulation in the disk state remains largely unexplored.

We find that the MF did not change significantly during the disk-to-pulsar state transition: from MF$_\textrm{disk}=26\pm10\%$ to MF$_\textrm{psr}=17\pm6\%$ (Section \ref{sec:J1227} and Figure ~\ref{fig:J1227_disk_vs_psr}).
The $0.1-1$ GeV gamma-ray flux from J1227 decreased by a factor $\simeq2.2$ during the transition, from $F_\textrm{disk}=(20.3\pm 0.9)\times10^{-12}\text{ erg cm}^{-2}\text{ s}^{-1}$ s$^{-1}$ in the disk state to $F_\textrm{psr}=(9.4\pm0.5)\times10^{-12}\text{ erg cm}^{-2}\text{ s}^{-1}$ in the pulsar state.
If the additional gamma-ray flux in the disk state were not modulated along the orbit, the MF would be lower in the disk state: it would drop by the same factor $\simeq2.2$, from the measured MF$_\textrm{psr} \simeq$17\% to an expected 8\%.
Instead, we find a somewhat higher fractional amplitude in the disk state (26\%).
Thus, our results suggest that the additional gamma-ray flux in the disk state is also orbitally modulated.

\subsection{Orbital inclination and eclipsing systems}

As outlined in Section~\ref{sec:intro}, the two main proposed mechanisms to explain GOM in spiders are synchrotron radiation of pulsar wind leptons in the companion's magnetic field \citep{vandermerwe20, clark21, sim24}, and inverse Compton scattering of blackbody photons from the companion by the energetic leptons from the pulsar \citep{an20, clark21, sim24}.
Here we argue that both scenarios predict a dependence of the MF on the orbital inclination ($i$), which is not seen in the data.

For six of our seven spiders with detected GOM, we found and compiled from the literature orbital inclinations $i$ determined from optical light curve modeling (Table \ref{tab:inclinations}). 
The five systems showing gamma-ray eclipses are especially interesting, since they allow us to probe GOM at the highest (edge-on) orbital inclinations.
Thus, we compiled the lower limits on their $i$ derived by \citealt{clark23} (also listed in Table \ref{tab:inclinations}).
The resulting MF-$i$ relation is shown in Figure~\ref{fig:MF_Inc}, where we find that edge-on systems do not feature stronger GOM.
Remarkably, we find MF$\lesssim$30\% for J1816 and B1957, despite them having $i\gtrsim80^\circ$.
The redback PSR~J1908+2105 is seen almost face on \citep[$i<6^\circ$,][]{Simpson25}, but it is relatively faint so our upper limits on MF are not constraining and we do not include it in Figure~\ref{fig:MF_Inc} (MF$<84\%$; cf. Tables~\ref{tab:properties} and \ref{tab:results}).
\begin{deluxetable}{llll}
\tablewidth{\textwidth}
\tabletypesize{\small}
\tablecaption{Orbital inclination estimates obtained from optical light curve modeling of spiders with detected GOM, and lower limits for gamma-ray eclipsing systems.
\label{tab:inclinations}}
\tablehead{
Name & Type & Inclination & Reference
}
\startdata
\textbf{J1124-3653} & BW & $44.9^{+4.0^\circ}_{-1.9}$ & \citealt{draghis19}\\
J1227-4853 & RBt & $46^\circ\lesssim i \lesssim 65^\circ$ & \citealt{demartino15}\\
J2039-5617 & RB & $60^\circ\lesssim i\lesssim78^\circ$ & \citealt{clark21} \\
\textbf{J2215+5135} & RB & $63.9^{+2.4^\circ}_{-2.7}$ & \citealt{linares18} \\
J2241-52364 & BW & $49.7^{+2.2^\circ}_{-1.9}$ & \citealt{draghis19}\\
J2339-0533 & RB & $69.3\pm2.3^\circ$ & \citealt{kandel19}\\
% \textbf{J1946-5403} & BW & Soft, pulsed & SC & SC [2] & * [7] \\
\hline
B1957+20 & BW & $>84.1^\circ$ & \citealt{clark23} \\
J1048+2339 & RB & $>80.4^\circ$ & \citealt{clark23} \\
J1555-2908 & BW & $>83.1^\circ$ & \citealt{clark23} \\
J1816+4510 & RB & $>79.0^\circ$ & \citealt{clark23} \\
J2129-0429 & RB & $>76.3^\circ$ & \citealt{clark23} \\
% J1311-3430 & BW & Hard, off-pulse & IC [17,18] & & 1 [6]\\
% J1702.7-5655 & RBc & Full & IC [19] & &  \\
% J1023+0038 ** & RBt & Hard & DN [20] & IC [21] & 1 [6]\\
\hline
\enddata
\tablecomments{RB: redback; RBt: redback and transitional MSP; BW: black widow.
}
% \tablenotetext{*}{text}
\end{deluxetable}

% J1124 and J2241 from \cite{draghis19} with  $44.9^{+4.0^\circ}_{-1.9} $ and $49.7^{2.2^\circ}_{-1.9}$; J2339 from \cite{kandel20} with $69.3\pm2.3^\circ$, J2215 from \cite{linares18} with  $63.9^{+2.4^\circ}_{-2.7}$; and J2039 from \cite{clark21} with $\sim75^\circ$, and a lower limit of $\gtrsim60^\circ$, and an upper limit of $\lesssim78^\circ$ based on the non-detection of gamma-ray eclipses.
% Additionally, we include inclination values derived from gamma-ray eclipse analysis by \cite{clark23} for PSR B1957+20, J1048+2339, J1555-2908, J1816+4510 and J2129-0429.
% The lower limits on the inclinations are $84.1^\circ$, $80.4^\circ$, $83.1^\circ$, $79.0^\circ$ and $76.3^\circ$, respectively.

In the inverse Compton scenario, electrons and positrons in the pulsar wind upscatter photons from the companion \citep{sim24}. 
The resulting flux is proportional to the seed photon energy density along the observer's line of sight.
If blackbody photons from the companion are emitted isotropically 
%and distributed 
with a number density that scales as $1/r^2$ (where $r$ is the radial distance from the center of the star), the line-of-sight seed photon energy density varies with the orbital phase.
This phase-dependent variation can produce GOM.
Furthermore, systems viewed edge-on, i.e., with high inclination, are expected to exhibit stronger modulation, as the line of sight passes closer to the companion, where photon densities are higher. 

In the synchrotron scenario, the modulation arises for two reasons.
First, the synchrotron peak energy is proportional to the strength of the companion's dipolar magnetic field, which decreases with distance as $1/r^3$. 
This peak falls within the LAT energy band near the pulsar's superior conjunction, when the energetic electrons pass closest to the companion, so the detected counts are higher\footnote{As shown by \cite{sim24}, this is true for for a companion magnetic field of $\sim 1$ kG and primary electrons with Lorentz factors $\sim10^7-10^8$.}.
At inferior conjunction, the peak shifts to lower energies since the magnetic field along the line of sight is lower, and the LAT band covers only the high-energy tail of the synchrotron spectrum (so the flux is lower).
Second, around pulsar inferior conjunction when the magnetic field along the line of sight is lowest, electrons radiate less efficiently as the synchrotron cooling time becomes longer than the residence time. During that phase, the synchrotron flux becomes proportional to $B^2$, giving the lowest gamma-ray fluxes \citep{sim24}.
In this synchrotron scenario, the magnetic field of the companion introduces an inclination dependence,  similar to the inverse Compton scenario.
As a result, edge-on systems are again expected to show stronger modulations.

\subsection{Modeling GOM in spiders}

We developed a simple analytical model to quantify the MF-$i$ relation in both scenarios.
We assume that both inverse Compton and synchrotron emissions are primarily generated at the point along the line of sight closest to the companion star, where pulsar wind electrons interact with the densest photon field and strongest magnetic field, respectively.
The distance from the center of the companion to this point is
\begin{equation}
R=a\sin\theta,
\label{eq:R}
\end{equation}
where $a$ is the binary separation, and $\theta$ is the angle between the line of sight and the line connecting the pulsar and its companion.
The angle $\theta$ depends on both the orbital phase $\Phi$ and the binary inclination $i$, and can be expressed as 
\begin{equation}
\cos\theta=\sin (i)\sin\left(2\pi\Phi\right).
\label{eq:theta}
\end{equation}

For the inverse Compton scenario, we use Equation (15) from \cite{sim24} describing the bolometric inverse Compton flux.
The $\Phi$ and $i$ dependence in the flux lies only in the companion's photon field energy density,
\begin{equation}
u_*=\frac{L}{4\pi R^2 c},
\label{eq:photon_field_1}
\end{equation}
where $L$ is the bolometric optical luminosity of the companion star.
Thus, we can write the inverse Compton flux as
\begin{equation}
F_\textrm{IC}=K_\textrm{IC}\times u_*(R),
\end{equation}
where $K_\textrm{IC}$ is a constant factor, independent of the binary phase and inclination, which characterizes the binary and pulsar wind properties \citep{sim24}.

\begin{figure}[ht!]
\includegraphics[width=0.47\textwidth]{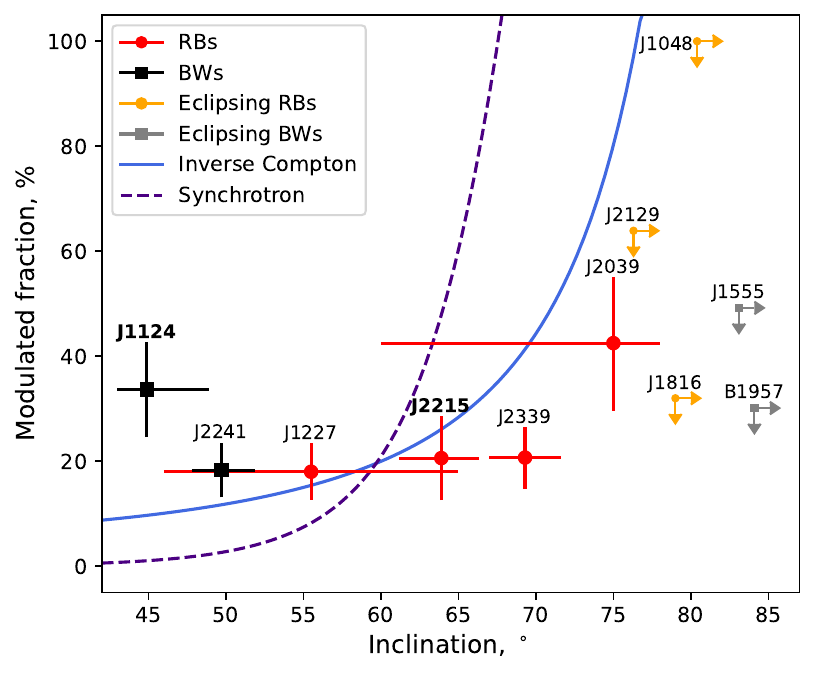}
\caption{Modulated fractions as a function of orbital inclination (red for redbacks, black for black widows). 
Upper/lower limits on MF/$i$ are shown for eclipsing systems, which have tightly constrained inclinations (orange for redbacks and gray for black widows).
Solid blue and dashed indigo lines show our predictions for the inverse Compton and synchrotron scenarios, respectively, with an arbitrarily chosen flux normalization.
\label{fig:MF_Inc}}
\end{figure}

Inserting \eqref{eq:R}, \eqref{eq:theta} into \eqref{eq:photon_field_1}, we obtain the photon field energy density  as a function of $\Phi$ and $i$:
\begin{equation}
u_*(\Phi, i)=\frac{L}{4\pi a^2c\left(1-\sin^2i\sin^2\left(2\pi\Phi\right)\right)}.
\end{equation}
% \cite{sim24} conclude that this model underestimates the modulated fluxes. 
Since $u_*$ carries the orbital phase and inclination dependence, and the MF is directly proportional to it at its maximum ($\Phi=0.25$), we fix the total gamma-ray flux normalization $F_\textrm{g}$ in order to estimate how the modulation strength depends on inclination.
Therefore, we calculate the MF as
\begin{equation}
\textrm{MF}_\textrm{IC}=\left(\frac{K_\textrm{IC}}{F_\textrm{g}}\right) \left( \frac{L}{4\pi a^2c}\right) \frac{1}{1-\sin^2 i}.
\end{equation}
The resulting $\textrm{MF}_\textrm{IC}-i$ relation is shown in Figure \ref{fig:MF_Inc} with a blue solid line.

To model the synchrotron case, we assume that electrons cool faster than their residence time (ensuring efficient synchrotron emission) and emit at the closest point to the companion, where the magnetic field is highest.
The synchrotron spectral energy distribution (SED) then shifts relative to the LAT band as $\Phi$ and $i$ change.
Since the number of emitting electrons is unknown, our simple model cannot predict absolute fluxes.
A full treatment would require integrating losses along the line of sight, as done by \cite{sim24}. 
Here, we focus on the relative changes of the SED.
Because the LAT band covers only the peak and the high-energy tail of the spectrum, contributions from weaker field regions, which affect mainly the low-energy tail, can be neglected.

We take a simplified approach in which the synchrotron emission is produced by monoenergetic electrons with Lorentz factor $\gamma_0$, propagating in the randomly oriented magnetic field of the companion, $B$ (\citealt{sim24}; see also \citealt{finke08}). 
Then, the flux density is
\begin{equation}
f_\textrm{SY}(E)=K_\textrm{SY}\times E\times \tilde{R}\left(\frac{E}{E_c}\right),
\label{eq:synch1}
\end{equation}
where $K_\textrm{SY}$ is a constant factor and
\begin{equation}
E_c=h\frac{3eB\gamma_0^2}{4\pi mc}
\end{equation}
is the characteristic (peak) energy of the synchrotron SED.
Here, $e$ and $m$ are the charge and mass of the electron,  $h$ is Planck's constant, $c$ is the speed of light and $B=B_0 (R_0/r)^3$ is the magnetic field in the companion region, where $B_0$ is the magnetic field strength at the stellar surface and $R_0$ the radius of the companion star.
Finally, $\tilde{R}$ is given by 
\begin{equation}
\tilde{R}(x)=\frac{x}{2}\int_0^\pi d\xi sin\xi \int _{x/sin\xi}^{\infty}dt K_{5/3}(t),
\end{equation}
where  $K_{5/3}$ is the modified Bessel function of the third kind.

Integrating Equation \eqref{eq:synch1} over the $0.1-1$ GeV band at superior conjunction ($\Phi=0.25$) for different inclinations, we obtain the dependence of the synchrotron modulation on binary inclination.
Fixing the total flux $F_\textrm{g}$, we calculate MF as
\begin{equation}
    \textrm{MF}_\textrm{SY}=\frac{1}{F_\textrm{g}}\int_{0.1\text{ GeV}}^{1\text{ GeV}}f_\textrm{SY}(E)dE.
\end{equation}

The resulting $\textrm{MF}_\textrm{SY}-i$ relation is shown in Figure \ref{fig:MF_Inc} with a dashed indigo line, compared with $\textrm{MF}_\textrm{IC}$
(blue solid line) and the measured MFs of the detected systems as a function of orbital inclination.

$MF_{\rm IC}$ and $MF_{\rm SY}$ are calculated by adopting typical values for the orbital separation ($a=10^{11}\text{ cm}$) and companion radius ($R_0=0.3R_\odot$, typical for redbacks).
Moreover, two fundamental parameters in the case of $MF_{\rm SY}$ are the magnetic field at the surface of the companion star $B_0$ and the Lorentz factor $\gamma_0$. 
We assume $\gamma_0=10^7$ and $B_0=1\text{ kG}$, as done by \cite{sim24}, since these values are able to reproduce the observed light curve of a few systems.
Despite the wide range of inclinations ($i\simeq45^\circ-90^\circ$), we find a lack of correlation between the measured MF and $i$.
This challenges the two main models proposed to explain GOM, since they predict stronger GOM in edge-on systems (see Figure~\ref{fig:MF_Inc}).

The absence of stronger gamma-ray modulation at high inclinations  
disfavors models involving the companion star, and suggests instead that the modulating region is located closer to the pulsar itself.
We propose that synchrotron radiation from energetic leptons in the pulsar wind interacting with the magnetic field of the IBS could be a possible mechanism. 
This would require a relatively strong magnetic field within the shock \citep[perhaps amplified by turbulence,][]{Bell04} and a short synchrotron cooling time. 
This would naturally explain orbital phasing and the inclination independence of our measured MF values.
Since we find MF$\simeq30\%$ in J1946 (Sec.~\ref{sec:J1946}), where the IBS is presumably wrapped around the companion, the modulation should occur in the central parts of the IBS where the shock tangent is always nearly perpendicular to the line of sight (for large orbital inclinations).
Future studies are required to quantify the modulated fraction in this or alternative models.
New detections of GOM or more stringent limits at lower orbital inclinations ($i<45^\circ$) may also give new clues as to its physical origin.

\section{Summary and Conclusions}

After conducting a systematic search for gamma-ray orbital modulation in 43 spider pulsars, we have discovered three new systems exhibiting soft pulsed GOM (and confirmed four previous detections).
We have quantified the modulated fraction in the $0.1-1$~GeV band, and found this to be approximately constant at MF=22.0$\pm$2.6~\%.
In all detected cases, the gamma-ray modulation peaks near the pulsar's superior conjunction, regardless of the maximum phase of the XOM.
In the transitional pulsar J1227, we find a tentative difference in the peak phase of the soft GOM between the disk and pulsar states, but the MFs remain approximately constant throughout the state transition.
We find that the measured MFs are relatively uniform and show no clear dependence on inclination.
This challenges current models proposed to explain GOM, including those based on synchrotron or inverse Compton emission, which predict a strong dependence of MF on orbital inclination.
Instead, we suggest that the region that modulates gamma-rays is closer to the pulsar (rather than the companion star), where the dependence on inclination can be weaker.

\section*{Acknowledgments}
MS is grateful to Marco Turchetta, Egor Podlesnyi, Bidisha Sen, Jordan Simpson, Karri Koljonen and Colin Clark for their help and useful discussions. 

This project has received funding from the European Research Council (ERC) under the European Union’s Horizon 2020 research and innovation programme (grant agreement No. 101002352, PI: M. Linares).

We acknowledge the use of data and timing information from the Fermi Large Area Telescope Third Pulsar Catalog (3PC; \citealt{3PC}).
% This research made use of data from the Fermi Large Area Telescope (Fermi-LAT) archive, provided by NASA's Fermi Science Support Center \footnote{\url{https://fermi.gsfc.nasa.gov/cgi-bin/ssc/LAT/LATDataQuery.cgi}}.

\facilities{Fermi-LAT}
\software{Fermi Science Tools (v2.2.0) \citep{2019ascl.soft05011F}; \texttt{fermipy} \citep{wood17}; \texttt{TEMPO2} \citep{tempo2, hobbs06, tempo2fermi}; NumPy \citep{harris2020array}; SciPy \citep{2020SciPy-NMeth}; Matplotlib \citep{Hunter:2007}}

% \bibliography{bib}{}
\bibliographystyle{aasjournal}

\appendix

\section{Folded orbital light curves}
\label{sec:appendixLCs}
\restartappendixnumbering

Figure \ref{fig:all_lcs} shows light curves folded at the orbital period for the 36 systems analyzed, with no significant GOM detected.

\begin{figure}[h!]
  \centering
  \includegraphics[width=0.83\textwidth]{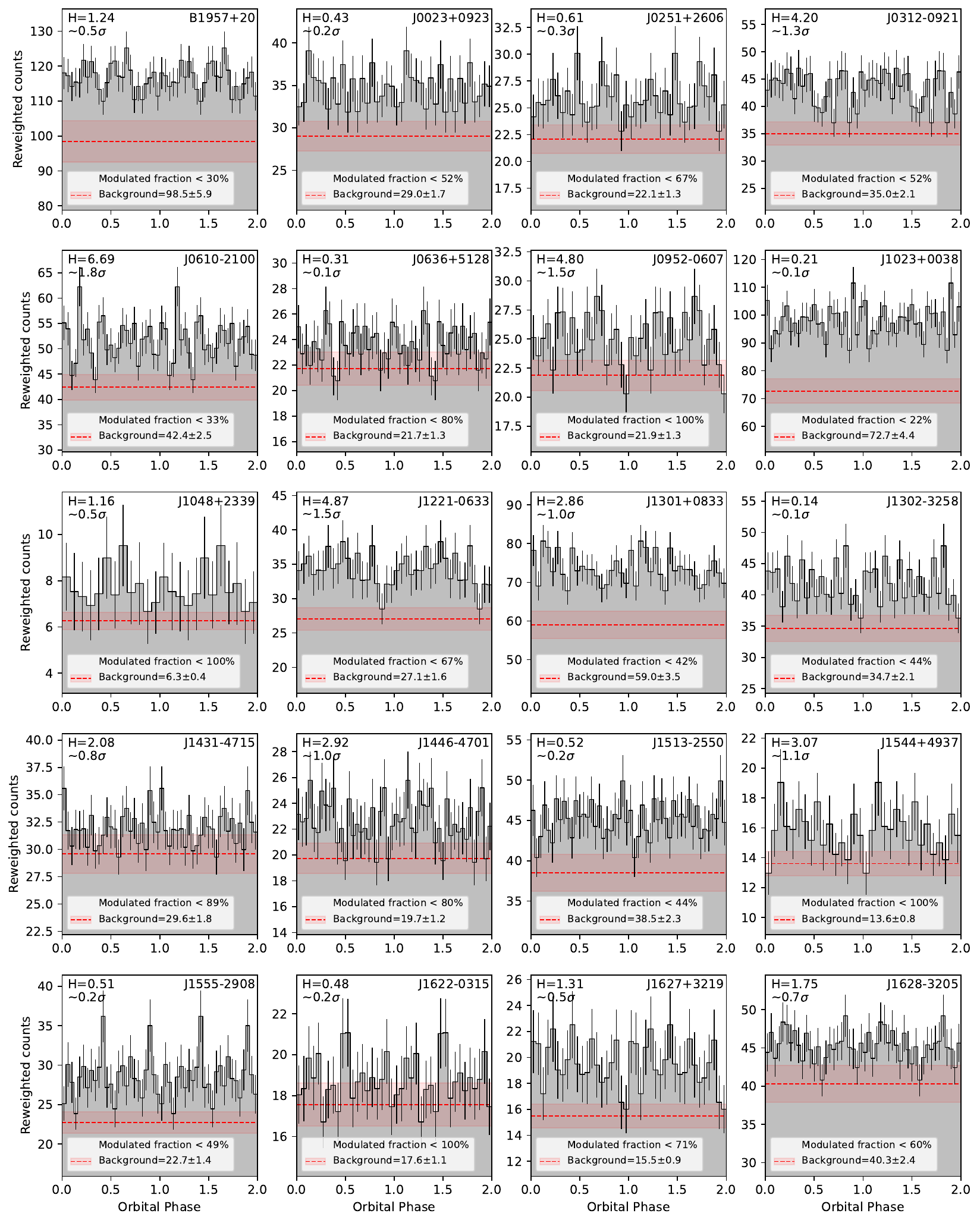 }
  \caption{Orbital light curves.}
  \label{fig:all_lcs}
\end{figure}

\begin{figure}[h!]
  \centering
  \includegraphics[width=0.83\textwidth]{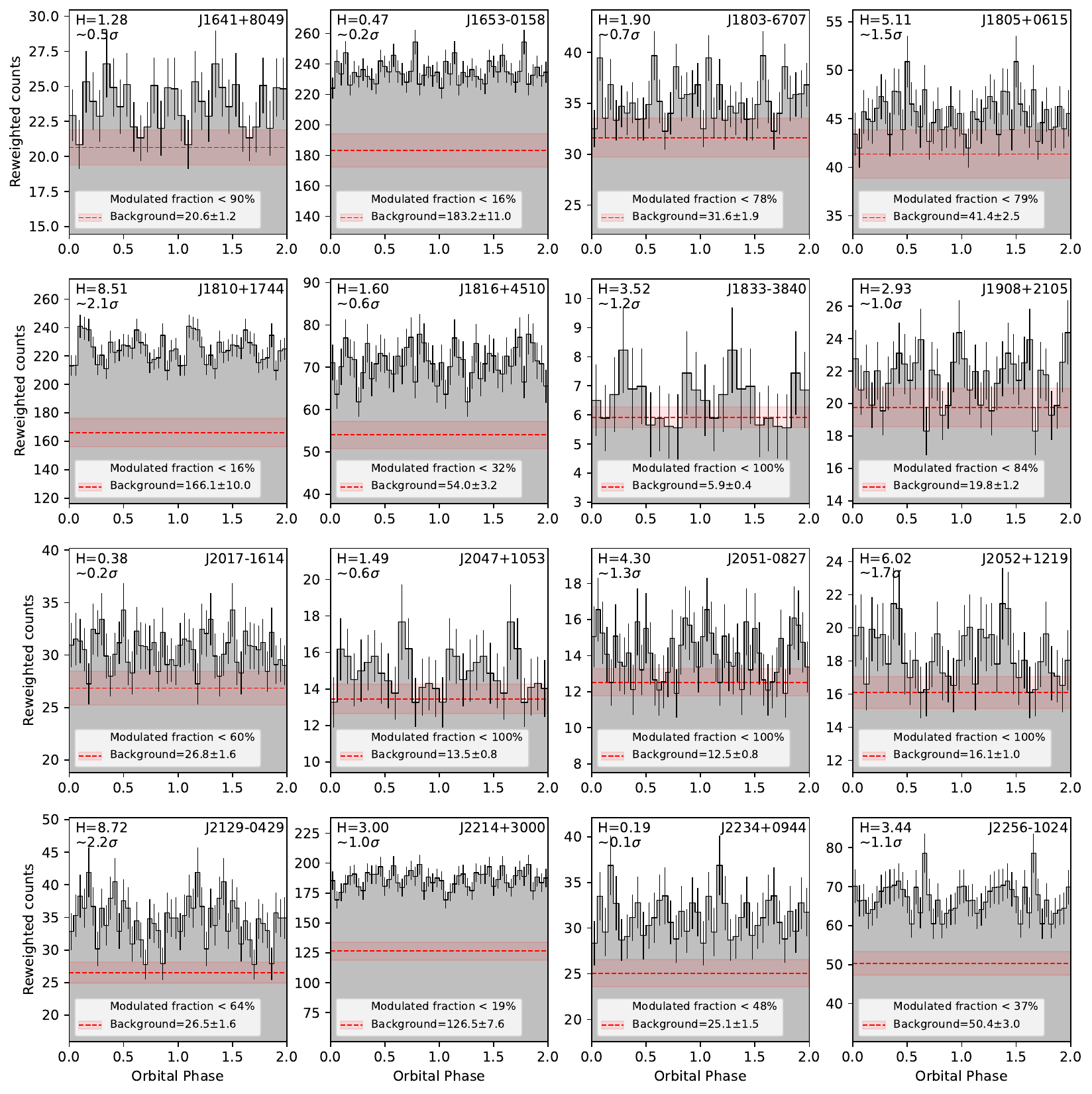 }
  \addtocounter{figure}{-1}% reuse same figure number
  \caption{(Continued) Orbital light curves.}
\end{figure}

\clearpage

\section{Extended timing products}
\restartappendixnumbering
\label{sec:exteded_timing}

\begin{figure*}[h]
\gridline{
  \fig{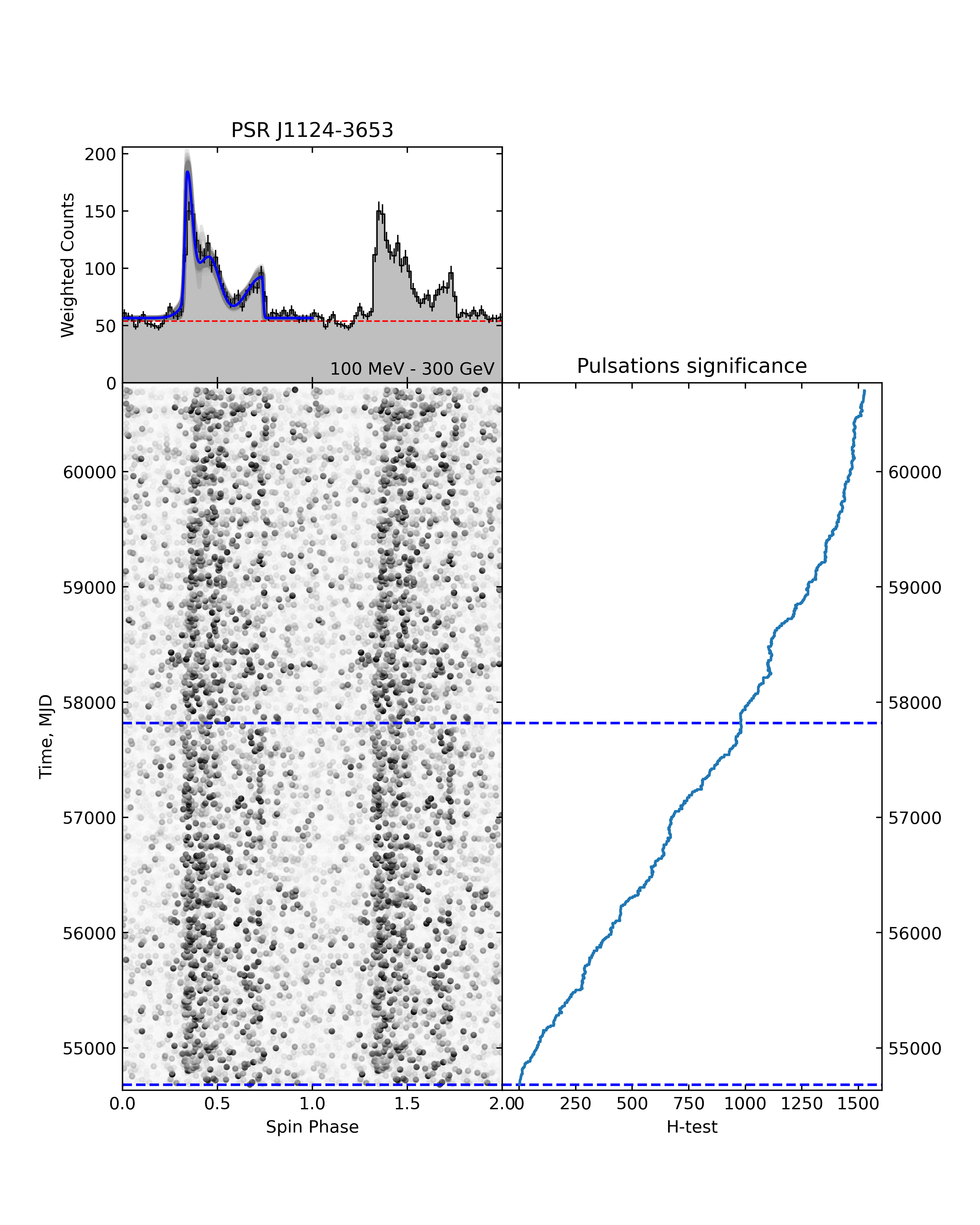}{0.48\textwidth}{}
  \fig{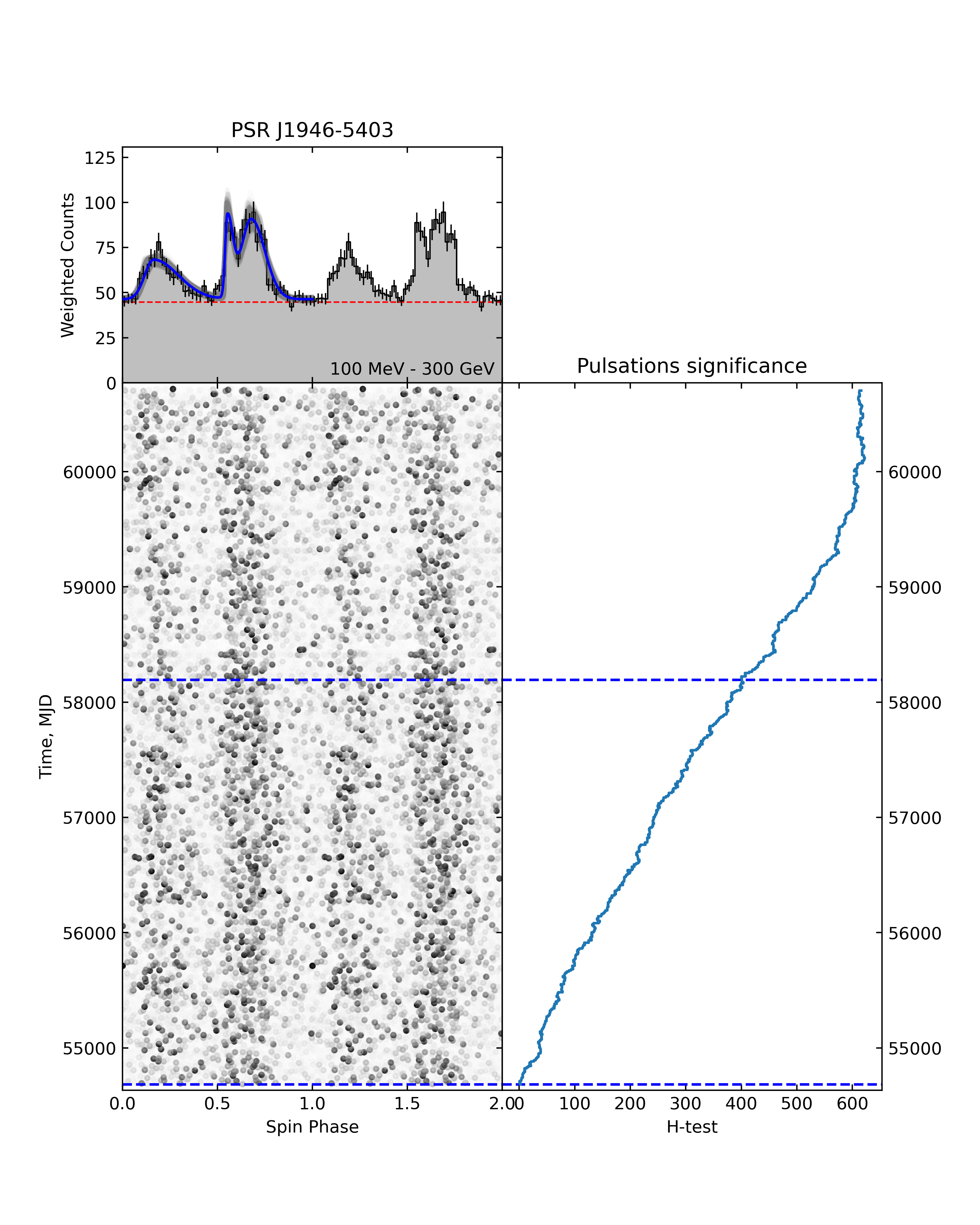}{0.48\textwidth}{}
}
\caption{Our gamma-ray timing results for J1124 and J1946 in the $0.1-300$ GeV energy band. 
\textit{Top panel}: the black curve represents the folded light curve, the blue curve shows the template pulse profile, and the dashed red line indicates the background level. 
\textit{Bottom left}: photon spin phases are plotted against time, with the color representing the weights of the photons. 
\textit{Bottom right}: the cumulative H-test for pulsations significance is shown over time. 
Dashed blue lines in the bottom panels indicate the start and end of the timing solution used.}
\label{fig:J1124_J1946_timing}
\end{figure*}

Figure \ref{fig:J1124_J1946_timing} shows  shows examples, where we extended 3PC timing solution.

\clearpage
\section{Folded light curves of eclipsing systems}
\restartappendixnumbering

\begin{figure*}[ht!]
\centering
\includegraphics[width=0.8\textwidth]{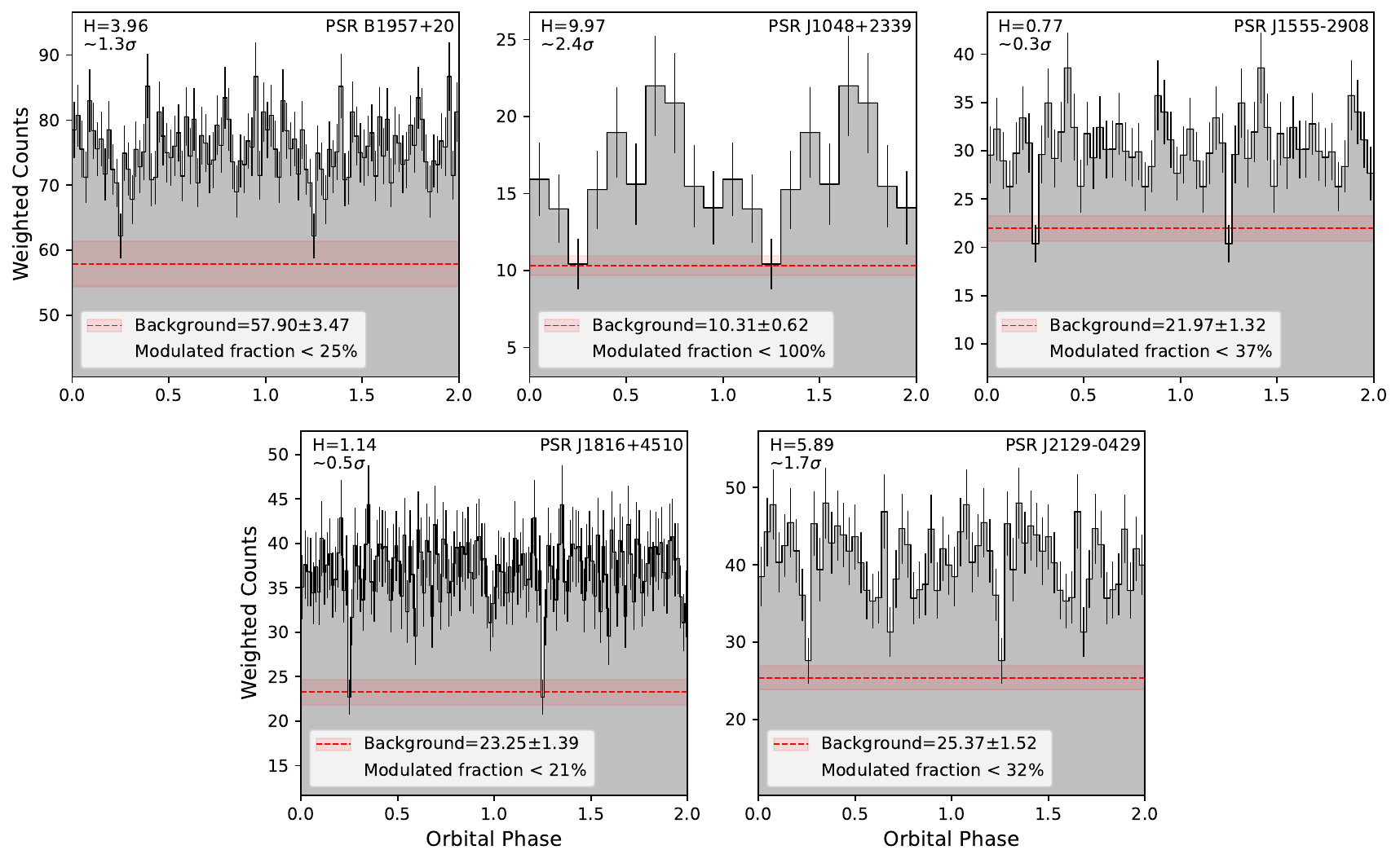}
\caption{Reweighted gamma-ray light curves of eclipsing systems in the 0.1 – 300 GeV energy range, optimal for eclipse detection. The black points represent the folded data, with uncertainties and the red line represents the background level.
\label{fig:eclipsing}}
\end{figure*}

\begin{figure*}[ht!]
\centering
\includegraphics[width=0.8\textwidth]{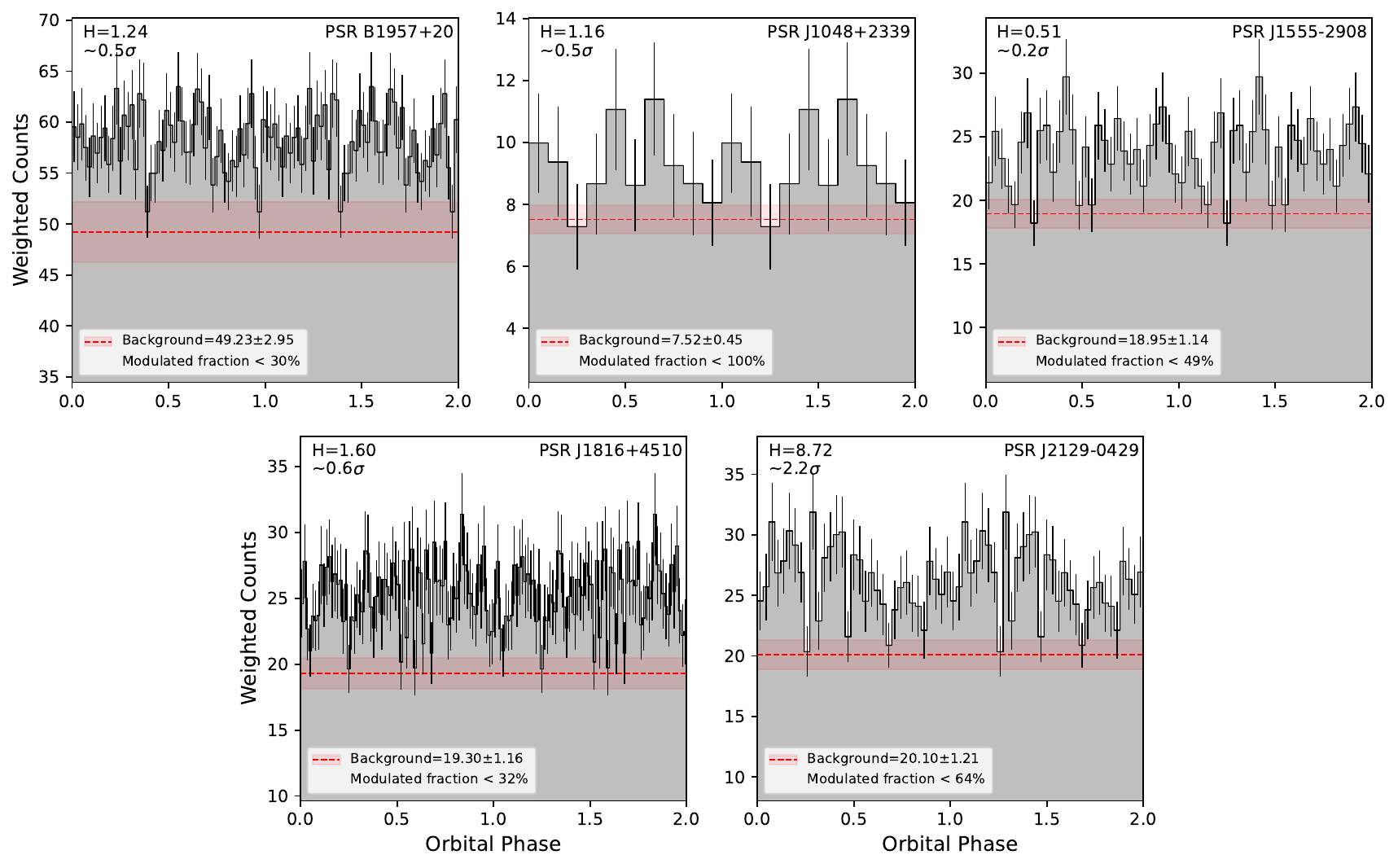}
\caption{Reweighted gamma-ray light curves of eclipsing systems in the 0.1 – 1 GeV energy range, optimal for GOM detection. The black points represent the folded data, with uncertainties and the red line represents the background level.
\label{fig:eclipsing_soft}}
\end{figure*}

Folded $0.1-300$ GeV and $0.1-1$ GeV light curves of eclipsing systems are shown in Figures \ref{fig:eclipsing} and \ref{fig:eclipsing_soft} respectively.  

\end{document}